\documentclass[sort&compress, 3p]{elsarticle}
\usepackage[utf8]{inputenc}
\usepackage{amsmath}
\usepackage{amssymb}
\usepackage{bm}
\usepackage{mathtools}
\usepackage{caption}
\usepackage{subcaption}
\usepackage{pdfpages}
\usepackage[capitalise]{cleveref}
\usepackage{siunitx}
\newtheorem{remark}{Remark}[section]
\newcommand{\deriv}[3][]{\ensuremath{\dfrac{\partial^{#1} {#2}}{\partial {#3}^{#1}}}}
\usepackage[linesnumbered, ruled]{algorithm2e}
\usepackage{algpseudocode}

\definecolor{mygreen}{RGB}{49,156,54}
\newcommand{\Div}[1][]{\ensuremath{\operatorname{Div} #1}}
\newcommand{\barbm}[1]{\ensuremath{\bar{\bm{#1}}}}
\newcommand{\hatmathbf}[1]{\ensuremath{\hat{\mathbf{#1}}}}
\newcommand{\hatbm}[1]{\ensuremath{\hat{\bm{#1}}}}

\newcommand{\sqbracket}[1]{\ensuremath{\left(#1\right)}}
\usepackage{floatrow}
\newfloatcommand{capbtabbox}{table}[][\FBwidth]
\definecolor{mygreen}{RGB}{0,0,0}

\begin{document}
\begin{frontmatter}
\title{A reduced order model for geometrically parameterized two-scale simulations of elasto-plastic microstructures under large deformations}
\author[1,3]{Theron Guo\corref{cor1}}\ead{t.guo@tue.nl}
\author[2,3]{Ond\v{r}ej Roko\v{s}}\ead{o.rokos@tue.nl}
\author[1,3]{Karen Veroy}\ead{k.p.veroy@tue.nl}
\cortext[cor1]{Corresponding author}
\address[1]{Centre for Analysis, Scientific Computing and Applications, Eindhoven University of Technology, 5612 AZ Eindhoven, The Netherlands}
\address[2]{Mechanics of Materials, Eindhoven University of Technology, 5612 AZ Eindhoven, The Netherlands}
\address[3]{Institute for Complex Molecular Systems, Eindhoven University of Technology, 5612 AZ Eindhoven, The Netherlands}
\begin{abstract}
In recent years, there has been a growing interest in understanding complex microstructures and their effect on macroscopic properties. In general, it is difficult to derive an effective constitutive law for such microstructures with reasonable accuracy and meaningful parameters. One numerical approach to bridge the scales is computational homogenization, in which a microscopic problem is solved at every macroscopic point, essentially replacing the effective constitutive model. Such approaches are, however, computationally expensive and typically infeasible in multi-query contexts such as optimization and material design. To render these analyses tractable, surrogate models that can accurately approximate and accelerate the microscopic problem over a large design space of shapes, material and loading parameters are required. In this work, we develop a reduced order model based on Proper Orthogonal Decomposition (POD), Empirical Cubature Method (ECM) and a geometrical transformation method with the following key features: (i)~large shape variations of the microstructure are captured, (ii)~only relatively small amounts of training data are necessary, and (iii)~highly non-linear history-dependent behaviors are treated. The proposed framework is tested and examined in two numerical examples, involving two scales and large geometrical variations. In both cases, high speed-ups and accuracies are achieved while observing good extrapolation behavior.
\end{abstract}
\end{frontmatter}

\section{Introduction}
\label{sec:introduction}

Driven by advances in additive manufacturing and tailorable effective properties of metamaterials, there has been a growing interest in understanding structure-property relationships of complex microstructures. These microstructures can typically be described by a few shape parameters, leading to distinct types of effective behavior. To investigate such structure-property relations and to find the optimal shape for a given application, simulations are often considered. These simulations are in general computationally expensive or even intractable for direct numerical simulation, especially for large-scale engineering applications, since considerably fine meshes are required to capture the complex microstructural geometry. By employing multi-scale methods {\color{mygreen}based on computational homogenization}~\cite{Geers2010,Kouznetsova2001} or domain decomposition methods~\cite{Farhat1991AAlgorithm}, such large-scale problems can be separated into many smaller subproblems, thus rendering them amenable for efficient numerical simulation.

{\color{mygreen}Domain Decomposition (DD) methods are particularly useful when the micro- and macroscale are of comparable size, i.e., in the absence of a clear scale separation. They can be categorized in overlapping and non-overlapping DD methods.
Regarding the latter, the domain is divided into subdomains and coupled at the interfaces. One notable method is the so-called FETI-DP~\cite{Farhat2001FETI-DP:Method}, where the problem is solved in the corner points of each subdomain and in the Lagrange multipliers that enforce the interface continuities. To increase the computational efficiency, model order reduction methods were combined with DD. In the Reduced Basis Element (RBE) method~\cite{Maday2002AMethod}, each subdomain is accelerated with a reduced basis and the interfaces are coupled weakly in a non-conforming manner with Lagrange multipliers. In~\cite{PhuongHuynh2013AEstimation}, the Static Condensation Reduced Basis Element (SCRBE) was introduced where the internal degrees of freedom of each subdomain are represented by a reduced basis and condensed out, resulting in a conforming approximation space on the interfaces (also referred to as ports in this context). Constructing optimal local approximation spaces for these ports in two-component systems was discussed in~\cite{Smetana2016OptimalProcedures}, and finding them by local solutions of the Partial Differential Equation (PDE) with random boundary conditions was proposed in~\cite{Buhr2018RandomizedReduction}. In the context of solid mechanics, recent applications of such methods include, for example,~\cite{Diercks2023MultiscaleReduction,Hernandez2020AECM-hyperreduction}. For a more comprehensive overview of concepts in localized model order reduction, the interested reader is referred to~\cite{Buhr2020LocalizedProblems}.}

If scale separation is assumed, i.e., when the length scale of the typical microstructural features is much smaller than that of the macrostructure, first-order computational homogenization can be employed. Here, the behavior of the microstructure dictates the (average) constitutive behavior of an effective macrostructural continuum model. By defining a Representative Volume Element (RVE) which models the fine-scale geometry of the microstructure in full detail, a coarse-grained representation of the macrostructure with a much coarser discretization can be assumed at the macroscale. At every macroscopic integration point, the macroscopic strain is used to specify a microscopic boundary value problem which, after solution, returns the effective stress and stiffness. Since a PDE needs to be solved at every macroscopic Gauss integration point, this methodology is still computationally expensive, and efficient ways for its solution are needed.

Several approaches to tackle this problem have been reported in the literature. {\color{mygreen}One class of methods are based on data-driven reduced order models. For these methods, the microscopic problem is solved several times to generate training data, and the data is subsequently used to learn an effective constitutive model. After training, the microscopic solver is not required anymore and replaced by the learned constitutive model. Several methods were proposed for elastic material models, see, e.g.,~\cite{Linka2021,Guo2021b,Le2015,Kirchdoerfer2016}, and also for history-dependent microstructures~\cite{Mozaffar2019,Wu2020a}. Even though highly efficient and accurate reduced order models can potentially be obtained, the extension of the methods~\cite{Linka2021,Guo2021b,Le2015,Kirchdoerfer2016} for inelastic materials is challenging, and the methods for history-dependent microstructures~\cite{Mozaffar2019,Wu2020a} often require vast amounts of data. To learn an elasto-plastic material model, Mozaffar et al.~\cite{Mozaffar2019} generated up to 15000 deformation paths each of 100 load steps. 9000 deformation paths of up to 2000 steps each were generated by Wu et al.~\cite{Wu2020a}.

Another class of methods attempts to accelerate the existing microscopic solver. For instance, if the Fast Fourier Transform (FFT)~\cite{Moulinec1998,Mishra2016} is used to simulate the microstructure, its solution can} be accelerated by the (nonuniform) Transformation Field Analysis (see, e.g.,~\cite{dvorak1992,Michel2003}), or Self-consistent Clustering Analysis~\cite{Liu2016,Yu2019}. One disadvantage of FFT is that geometrical parameterizations of the RVE cannot be directly treated and, hence, sensitivities for material optimization cannot be directly computed. {\color{mygreen}If the microscopic problem is solved via the Finite Element (FE) method, the resulting multi-scale formulation is referred to as FE$^2$~\cite{Feyel1999,Matous2017,Miehe2002}. By directly solving the microscopic PDE with FE}, material or shape parameterizations can be considered in a straightforward manner, making the approach more suitable for inverse problems and optimization. To speed up the microstructural simulation, Proper Orthogonal Decomposition (POD)~\cite{Quarteroni2015,Hesthaven2016} can be utilized to find a reduced set of basis functions; the method then computes the Galerkin projection of the solution onto the space spanned by the snapshots. Although POD generally requires many full-order solves for training, it typically works well for all input parameters. In the context of first-order homogenization, POD was first applied in Yvonnet et al.~\cite{Yvonnet2007} for a hyper-elastic RVE, and later explored in Radermacher et al.~\cite{Radermacher2016a} for an elasto-plastic RVE under small strains. However, due to the non-linearities of the microscopic problem, the speed-ups were limited since the global force vector and stiffness matrix must be assembled by full integration in every microscopic Newton iteration. To address this issue, a further reduction called hyperreduction is required, which aims at finding an efficient way of assembling microstructural force and stiffness quantities. Notable hyperreduction methods are Empirical Interpolation Method (EIM) (see, e.g.,{\color{mygreen}~\cite{Barrault2004}}), a variant of EIM called Discrete Empirical Interpolation Method (DEIM) (see, e.g.,~\cite{Chaturantabut2010}), energy-based mesh sampling and weighting~\cite{Farhat2014DimensionalEfficiency}, reduced integration domain~\cite{Ryckelynck2009}, empirical quadrature procedure~\cite{Yano2019AnPDEs}, and Empirical Cubature Method (ECM)~\cite{Hernandez2017}. EIM and DEIM interpolate the non-linear integrand of the global force vector such that the integrals can be pre-computed. In~\cite{Hernandez2014,Soldner2017}, DEIM was used successfully to accelerate the solution of the microscopic PDE. However, these works only discussed the solution of the microscopic PDE and did not derive the effective stress and stiffness quantities required for the macroscopic problem. A possible disadvantage of EIM and DEIM is that they lead to non-symmetric tangent matrices, which might result in convergence issues, observed in, for instance,~\cite{Soldner2017,Radermacher2016}. The {\color{mygreen}rest} of the above-mentioned hyperreduction methods aim at approximating the integrals by finding a subset of integration points with corresponding {\color{mygreen} positive} weights among the set of all integration points used in the formulation of the microstructural PDE. This has the advantage that the stiffness matrix is always symmetric {\color{mygreen} and at least positive semi-definite (in practice usually positive definite unless instabilities occur)}, ensuring a good convergence of the microscopic problem. Example applications of hyperreduction methods have been successfully employed in two-scale simulations in~\cite{Caicedo2019HighModeling}, where an elasto-plastic composite RVE under large deformations was considered. In~\cite{Raschi2021}, a damage model for a composite RVE under small deformations was shown. While both works obtained accurate results and successfully accelerated the forward simulations of a two-scale problem, such formulations were limited to fixed microstructures only, i.e., did not account for possible parameterizations. In order to allow for optimization of microstructures, the surrogate model needs to be extended to a wide range of different design parameters (including geometrical as well as material).

This work aims to address this gap, by developing a hyper-reduced surrogate model for geometrically parameterized microstructures, to enable (shape) sensitivity analysis and optimization of materials. Furthermore, we intend to provide a detailed analysis of the reduced RVE problem for arbitrary loading paths and geometries and elucidate possible issues due to reduction. Our main contributions are:
\begin{enumerate}
    \item development of a hyper-reduced POD model for a family of geometrically parameterized microstructures, by employing a geometrical transformation method~\cite{Guo2022LearningParameterizations} and by extending the ECM algorithm to geometrical parameters,
    \item consistent derivation of the effective stress and stiffness of the hyper-reduced model,
    \item an empirical analysis of the accuracy of the surrogate model for elasto-plastic RVEs under large deformations for different geometries and loading conditions,
    \item a quantitative comparison of a two-scale example with continuous change in microstructural heterogeneities.
\end{enumerate}

The remainder of this paper is organized as follows. In~\cref{sec:problem}, the microscopic problem arising in first-order computational homogenization is briefly summarized, together with the computation of the effective quantities.~\cref{sec:surrogate} covers in-depth development of the reduced order model with particular focus on the empirical cubature method for geometrically parameterized microstructures, and includes a detailed derivation of the effective stress and stiffness. In~\cref{sec:results}, the proposed method is examined and tested in detail, first for a single RVE and then also for a full two-scale problem. Finally, a summary on the findings and concluding remarks are given in~\cref{sec:conclusion}.

In this work, the following notational convention is adopted. Italic bold symbols are used for coordinates $\bm{X}$ and vectorial or tensorial fields, such as the displacement field $\bm{u}$ or stress field $\bm{P}$. Upright bold symbols are used for algebraic vectors and matrices, such as the global stiffness matrix $\mathbf{K}$ or the coefficients of the discretized displacement field $\mathbf{u}$. A field quantity $\bm{u}$ for given parameters $\bm{\mu}$ is denoted as $\bm{u}(\bm{X}; \bm{\mu})$. Given second-order tensors $\bm{A}$ and $\bm{B}$, fourth-order tensor $\bm{C}$, and vector $\bm{v}$, the following operations are used: {\color{mygreen}$(\bm{A}\bm{B})_{ik} = A_{ij}B_{jk}$}, $\bm{A}:\bm{B} = A_{ij}B_{ij}$, $\bm{A}:\bm{C}:\bm{B} = A_{ij}C_{ijkl}B_{kl}$ and {\color{mygreen}$(\bm{A}\bm{v})_i = A_{ij}v_j$}, where the Einstein summation convention is implied.

\section{Formulation Of The Microscopic Problem}
\label{sec:problem}
In multiscale schemes based on first-order homogenization, {\color{mygreen}the macroscopic problem is governed by the standard linear momentum balance, but the macroscopic constitutive model (relating strains to stresses and stiffness)} is replaced by a microscopic PDE {\color{mygreen}(again governed by the standard linear momentum balance)} which is defined on an RVE{\color{mygreen}}. By prescribing the macroscopic deformation gradient on the microscale, the PDE can be solved and an effective stress and stiffness returned to the macroscopic solver, see~\cref{fig:geometric_parameterization}. For applications such as microstructure optimization, it is reasonable to additionally introduce a parameterization of the RVE in order to compute the sensitivities with respect to design variables. The microscopic boundary value problem is formulated below on a parameterized domain, as is usually the case in shape optimization. For brevity, the dependence on the macroscopic coordinates is omitted and a fixed macroscopic material point is assumed unless otherwise specified.

\subsection{Boundary Value Problem}
\begin{figure}[t]
    \centering
    \includegraphics[width=0.8\textwidth]{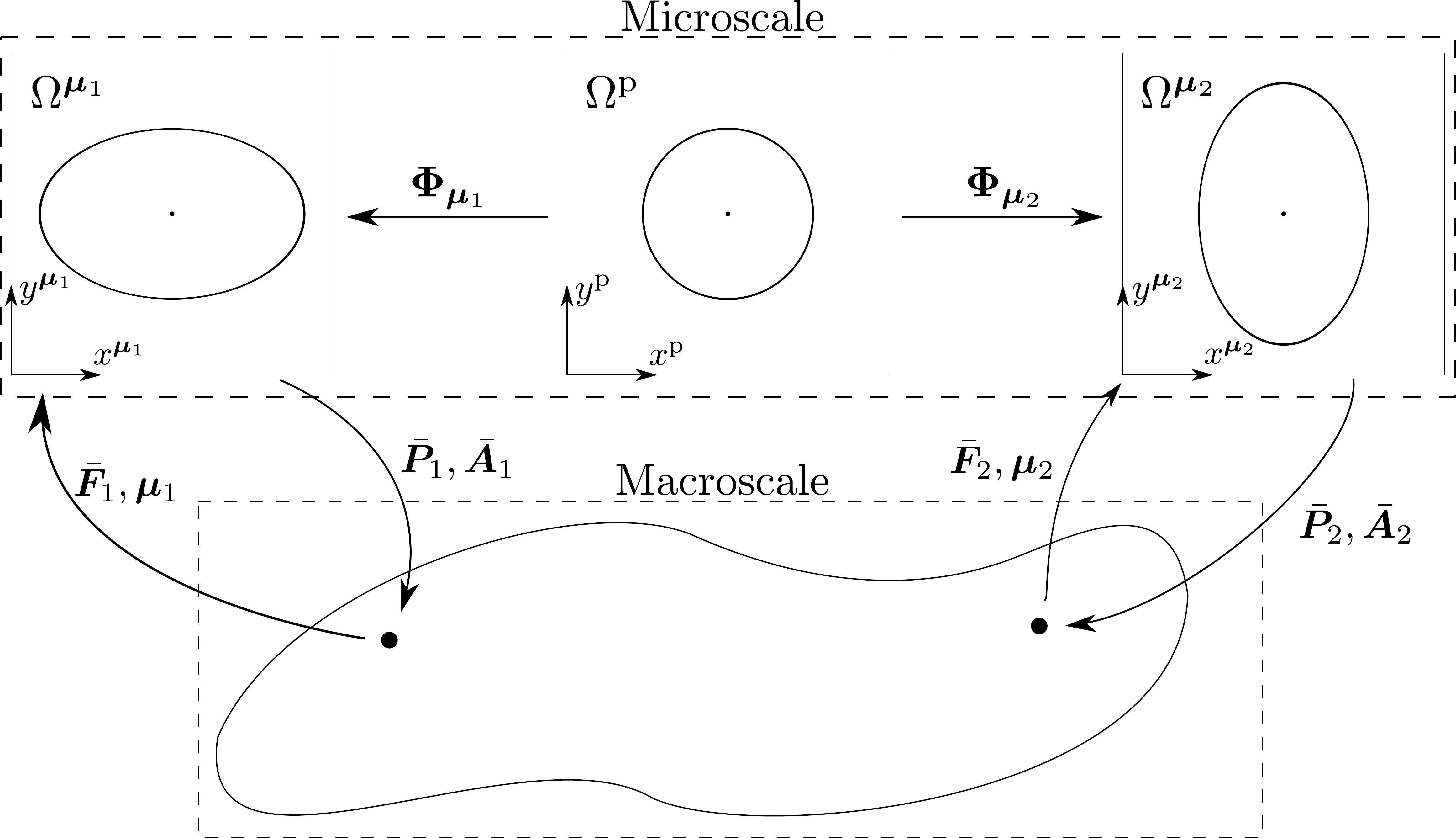}
    \caption{Two-scale problem based on first-order homogenization. At every macroscopic point, a microscopic simulation is defined through deformation gradient $\barbm{F}$ and shape parameters $\bm{\mu}$, and solved to obtain an effective stress $\barbm{P}$ and stiffness $\barbm{A}$. For different macroscopic points, different parameterized microstructures can be considered through $\bm{\mu}$. As an example of a family of geometrically parameterized microstructures, a parent domain with a circular inclusion $\Omega^{\rm{p}}$ (center), can be mapped onto parameterized domains $\Omega^{\bm{\mu}_1}$ (left) and $\Omega^{\bm{\mu}_2}$ (right) with mapping $\bm{\Phi}_{\bm{\mu}_1}$ and $\bm{\Phi}_{\bm{\mu}_2}$.}
    \label{fig:geometric_parameterization}
\end{figure}
Consider a family of domains $\Omega^{\bm{\mu}} \subset \mathbb{R}^{d}$ with space dimension $d=2,3$, parameterized by geometrical parameters $\bm{\mu}$, and spanned by a position vector $\bm{X}^{\bm{\mu}}\in\Omega^{\bm{\mu}}$. In \cref{fig:geometric_parameterization}, an example parent domain with a circular inclusion $\Omega^{\rm{p}}$ is geometrically parameterized and mapped to two distinct parameterized domains with elliptical inclusions, $\Omega^{\bm{\mu}_1}$ and $\Omega^{\bm{\mu}_2}$. The volume and the topology of the domain $|\Omega^{\bm{\mu}}|$ are assumed to remain fixed for all parameters (the outer boundaries of the RVE domain are fixed while the shape of the interior geometry can change). With the assumption of scale separation between macro- and microscale, the microscopic displacement field on the parameterized domain $\bm{u}(\bm{X}^{\bm{\mu}})$ can be written as the summation of a mean field $\barbm{u}(\bm{X}^{\bm{\mu}})$ and a fluctuation field $\bm{w}(\bm{X}^{\bm{\mu}})$, i.e., $\bm{u}(\bm{X}^{\bm{\mu}}) = \barbm{u}(\bm{X}^{\bm{\mu}}) + \bm{w}(\bm{X}^{\bm{\mu}})$. The mean field is fully specified through $\barbm{u}(\bm{X}^{\bm{\mu}}) \coloneqq (\barbm{F}-\bm{I})\bm{X}^{\bm{\mu}}$, where $\barbm{F}$ is the macroscopic deformation gradient tensor and $\bm{I}$ is the identity tensor. The total deformation gradient tensor $\bm{F}$ is defined as
\begin{align}
    \bm{F}(\bm{w}(\bm{X}^{\bm{\mu}})) \coloneqq \bm{I} + \dfrac{\partial \bm{u}}{\partial \bm{X}^{\bm{\mu}}} = \barbm{F} + \dfrac{\partial \bm{w}}{\partial \bm{X}^{\bm{\mu}}}.
\end{align}
The governing microscopic PDE is given as
\begin{alignat}{2}
\begin{aligned}
    \Div \bm{P}^T(\bm{F}(\bm{w}(\bm{X}^{\bm{\mu}}))) &= \bm{0}         &&\text{ on } \Omega^{\bm{\mu}}, \\
    \bm{w} &\text{ periodic } \quad &&\text{ on } \partial\Omega^{\bm{\mu}},
\end{aligned}
\end{alignat}
where $\Div(\bullet)$ is the divergence operator with respect to $\bm{X}^{\bm{\mu}}$ and $\bm{P}$ denotes the second-order first Piola-Kirchhoff (1PK) stress tensor. No constitutive model is specified at this point, although we assume that the stress $\bm{P}$ is a non-linear function of the deformation gradient $\bm{F}$ (or its history). The weak form of the problem is then: given the macroscopic deformation gradient $\barbm{F}$, find the fluctuation field $\bm{w}^* \in \mathcal{V}\coloneqq\{\bm{v} \in (H^1(\Omega^{\bm{\mu}}))^d \ | \ \bm{v} \text{ periodic on } \partial\Omega^{\bm{\mu}}\}$ that fulfills
\begin{align}
G(\bm{w}, {\color{mygreen}\delta\bm{w}}) \coloneqq \int_{\Omega^{\bm{\mu}}} \frac{\partial \delta\bm{w}}{\partial \bm{X}^{\bm{\mu}}} : \bm{P}\sqbracket{\barbm{F} + \frac{\partial \bm{w}}{\partial \bm{X}^{\bm{\mu}}}} d\bm{X}^{\bm{\mu}} \stackrel{!}{=} 0, \qquad \forall \delta\bm{w}\in\mathcal{V},
\label{eq:weakform-dep}
\end{align}
where the integral bounds depend on the parameters $\bm{\mu}$, $\delta\bm{w}$ denotes a test function, and $H^1(\Omega^{\bm{\mu}})$ is a Hilbert space with square integrable functions and square integrable derivatives. The inner product in $\mathcal{V}$ is defined as
\begin{align}
    (\bm{u}, \bm{v})_\mathcal{V} \coloneqq \int_{\Omega^{\bm{\mu}}} \left(\bm{u}\cdot \bm{v} + \frac{\partial \bm{u}}{\partial \bm{X}^{\bm{\mu}}} : \frac{\partial \bm{v}}{\partial \bm{X}^{\bm{\mu}}}\right) d\bm{X}^{\bm{\mu}}. \label{eq:Vdotproduct}
\end{align}
From~\cref{eq:weakform-dep}, it is apparent that the macroscopic deformation gradient $\barbm{F}$ represents the external loading, while the fluctuation displacement field $\bm{w}$ balances the system. To simplify the problem in~\cref{eq:weakform-dep} and remove the parameter dependence of the integral bounds, a parent domain $\Omega^{\rm{p}}$ is defined. To this end, we assume that there exists a parameter-dependent diffeomorphism $\bm{\Phi}_{\bm{\mu}} : \Omega^{\rm{p}} \rightarrow \Omega^{\bm{\mu}}, \bm{X}^{\rm{p}} \mapsto \bm{X}^{\bm{\mu}}$, see~\cref{fig:geometric_parameterization}. Using integration by substitution, the problem of~\cref{eq:weakform-dep} can be restated as follows: given the macroscopic deformation gradient $\barbm{F}$, find $\bm{w}^{*\rm{p}} \in \mathcal{V}^{\rm{p}}\coloneqq\{\bm{v} \in (H^1(\Omega^{\rm{p}}))^d \ | \ \bm{v} \text{ periodic on } \partial\Omega^{\rm{p}}\}$ that fulfills
\begin{align}
G^{\rm{p}}(\bm{w}^{\rm{p}}, {\color{mygreen}\delta\bm{w}^{\rm{p}}}) \coloneqq \int_{\Omega^{\rm{p}}} \left(\frac{\partial \delta\bm{w}^{\rm{p}}}{\partial \bm{X}^{\rm{p}}} \bm{F}_{\bm{\mu}}^{-1}\right) : \bm{P}\sqbracket{\barbm{F} + \frac{\partial \bm{w}^{\rm{p}}}{\partial \bm{X}^{\rm{p}}} \bm{F}_{\bm{\mu}}^{-1}} \left|\det \bm{F}_{\bm{\mu}}\right| d\bm{X}^{\rm{p}} = 0, \qquad \forall \delta\bm{w}^{\rm{p}}\in\mathcal{V}^{\rm{p}}, \label{eq:weakform-dep-mapped}
\end{align}
with the transformation gradient $\bm{F}_{\bm{\mu}} \coloneqq \dfrac{\partial \bm{\Phi_{\bm{\mu}}}}{\partial \bm{X}^{\rm{p}}}$ and $d\bm{X}^{\bm{\mu}} = \left|\det \bm{F}_{\bm{\mu}}\right| d\bm{X}^{\rm{p}}$. The superscript p is used to denote quantities pertinent to the parent domain, e.g., $\bm{w}(\bm{X}^{\bm{\mu}}) = (\bm{w}\circ \bm{\Phi}_{\bm{\mu}})(\bm{X}^{\rm{p}})=\bm{w}^{\rm{p}}(\bm{X}^{\rm{p}})$. To iteratively solve the non-linear problem in~\cref{eq:weakform-dep-mapped}, a linearization using the Gateaux derivative around the current state $\bm{w}^{\rm{p}}$ in direction $\Delta \bm{w}^{\rm{p}} \in \mathcal{V}^{\rm{p}}$ is required and can be written as,
\begin{align}
\begin{aligned}
\left.\frac{\partial G^{\rm{p}}(\bm{w}^{\rm{p}} + \tau \Delta \bm{w}^{\rm{p}}, {\color{mygreen}\delta\bm{w}^{\rm{p}}})}{\partial \tau}\right|_{\tau=0} &= \int_{\Omega^{\rm{p}}} \left(\frac{\partial \delta\bm{w}^{\rm{p}}}{\partial \bm{X}^{\rm{p}}} \bm{F}_{\bm{\mu}}^{-1} \right) : \bm{A}\sqbracket{\barbm{F} + \frac{\partial \bm{w}^{\rm{p}}}{\partial \bm{X}^{\rm{p}}} \bm{F}_{\bm{\mu}}^{-1}} : \left(\frac{\partial \Delta\bm{w}^{\rm{p}}}{\partial \bm{X}^{\rm{p}}} \bm{F}_{\bm{\mu}}^{-1}\right) \left|\det \bm{F}_{\bm{\mu}}\right| d\bm{X}^{\rm{p}},
\end{aligned} \label{eq:Dweakform-dep}
\end{align}
where $\bm{A} \coloneqq \dfrac{\partial \bm{P}}{\partial \bm{F}}$ is the fourth-order stiffness tensor. Once the transformation map $\bm{\Phi}_{\bm{\mu}}$ is known,~\cref{eq:weakform-dep} can be solved on the parent domain using~\cref{eq:weakform-dep-mapped,eq:Dweakform-dep}. Further details on how to find these transformations for a range of geometrical parameters are provided in~\cref{subsec:auxiliary}.

By {\color{mygreen}following a standard Galerkin} finite element discretization for $\bm{w}^{\rm{p}}\approx \bm{w}^{\rm{p}}_h\in \mathcal{V}_h^{\rm{p}} \subset \mathcal{V}^{\rm{p}}$, with $\dim\mathcal{V}_h^{\rm{p}}=\mathcal{N}$, the number of degrees of freedom of the dicretization, the internal force vector $\mathbf{f}\in\mathbb{R}^{\mathcal{N}}$ and global stiffness matrix $\mathbf{K}\in\mathbb{R}^{\mathcal{N}\times\mathcal{N}}$ can be derived from~\cref{eq:weakform-dep-mapped,eq:Dweakform-dep}, resulting in the following non-linear system of equations
\begin{align}
    \mathbf{f}(\mathbf{w}) = \bm{0}, \label{eq:full_system}
\end{align}
where $\mathbf{w}\in\mathbb{R}^{\mathcal{N}}$ is the column vector of unknown coefficients of the discretized fluctuation field. This problem can be solved with the Newton method, i.e.,
\begin{align}
\begin{aligned}
\mathbf{K}(\mathbf{w}^m) \Delta\mathbf{w} &= -\mathbf{f}(\mathbf{w}^m), \\
    \mathbf{w}^{m+1} &= \mathbf{w}^m + \Delta\mathbf{w},
\end{aligned} \label{eq:full_newton}
\end{align}
where $m$ is the Newton iteration number and \cref{eq:full_newton} is repeated until $||\mathbf{f}(\mathbf{w}^m)||_2 \leq \varepsilon_{\text{newton}}$ with $\varepsilon_{\text{newton}}$ a user-defined tolerance. {\color{mygreen}For more details on the finite element method and discretization of weak forms, we refer to~\cite{Belytschko2014NonlinearStructures}.}

\subsection{Effective Quantities}
\label{subsec:eff_quantities_full}
For conciseness of notation, the following abbreviations are introduced to denote quantities after the solution $\bm{w}^{*\rm{p}}$ has been obtained:
\begin{align}
    \bm{P}^{*\rm{p}} &\coloneqq \bm{P}\sqbracket{\barbm{F} + \frac{\partial \bm{w}^{*\rm{p}}}{\partial \bm{X}^{\rm{p}}} \bm{F}_{\bm{\mu}}^{-1}}, \label{eq:converged_stress}\\
    \bm{A}^{*\rm{p}} &\coloneqq \bm{A}\sqbracket{\barbm{F} + \frac{\partial \bm{w}^{*\rm{p}}}{\partial \bm{X}^{\rm{p}}} \bm{F}_{\bm{\mu}}^{-1}}. \label{eq:converged_stiffness}
\end{align}
Upon obtaining solution $\bm{w}^{*\rm{p}}$ from~\cref{eq:weakform-dep-mapped}, the effective stress is computed as
\begin{align}
    \barbm{P} &\coloneqq |\Omega^{\rm{p}}|^{-1}\int_{\Omega^{\rm{p}}} \bm{P}^{*\rm{p}} |\det \bm{F}_{\bm{\mu}}| d\bm{X}^{\rm{p}}, \label{eq:eff_stress_full}
\end{align}
and the effective stiffness (in index notation) as
\begin{align}
\begin{aligned}
    \bar{A}_{ijkl} \coloneqq &\frac{\partial \bar{P}_{ij}}{\partial \bar{F}_{kl}} \\
    = &|\Omega^{\rm{p}}|^{-1} \frac{\partial}{\partial \bar{F}_{kl}} \int_{\Omega^{\rm{p}}} P^{*\rm{p}}_{ij} |\det \bm{F}_{\bm{\mu}}| d\bm{X}^{\rm{p}} \\
    = &|\Omega^{\rm{p}}|^{-1} \int_{\Omega^{\rm{p}}} A^{*\rm{p}}_{ijmn} \left(\mathbb{I}_{mnkl} + \frac{\partial}{\partial \bar{F}_{kl}}\left(\frac{\partial w_m^{*\text{p}}}{\partial X_r}\right)\left(F_{\bm{\mu}}^{-1}\right)_{rn}\right) |\det \bm{F}_{\bm{\mu}}| d\bm{X}^{\rm{p}},
\end{aligned} \label{eq:eff_stiffness_full}
\end{align}
where $\mathbb{I}_{mnkl} \coloneqq \delta_{mk}\delta_{nl}$ is the fourth-order identity tensor. To determine $\dfrac{\partial}{\partial \bar{F}_{kl}}\left(\dfrac{\partial w_m^{*\text{p}}}{\partial X_r}\right)$,~\cref{eq:weakform-dep-mapped} is differentiated with respect to $\barbm{F}$. For one particular component $\bar{F}_{kl}$ (where the indices $k$ and $l$ are assumed to be temporarily fixed), the differentiation yields
\begin{align}
\int_{\Omega^{\rm{p}}} \left(\frac{\partial \delta\bm{w}^{\rm{p}}}{\partial \bm{X}^{\rm{p}}} \bm{F}_{\bm{\mu}}^{-1}\right) : \bm{A}^{*\rm{p}} : \left(\frac{\partial \bm{q}_{kl}}{\partial \bm{X}^{\rm{p}}} \bm{F}_{\bm{\mu}}^{-1}\right) \left|\det \bm{F}_{\bm{\mu}}\right| d\bm{X}^{\rm{p}} = -\left(\int_{\Omega^{\rm{p}}} \left(\frac{\partial \delta\bm{w}^{\rm{p}}}{\partial \bm{X}^{\rm{p}}} \bm{F}_{\bm{\mu}}^{-1}\right) : \bm{A}^{*\rm{p}} \left|\det \bm{F}_{\bm{\mu}}\right| d\bm{X}^{\rm{p}}\right) : \mathbb{E}_{kl},\label{eq:tangentproblem}
\end{align}
where a new auxiliary vector field $\bm{q}_{kl} \coloneqq \dfrac{\partial \bm{w}^{*\rm{p}}}{\partial \bar{F}_{kl}} \in \mathcal{V}^{\rm{p}}$ has been defined (reflecting the sensitivity of the microfluctuation field with respect to the change of the applied macroscopic loading), and $\mathbb{E}_{kl} \in \mathbb{R}^{d\times d}$ is a second order tensor with all entries zero, except for the $kl$-th entry which is 1. The linear tangent problem of \cref{eq:tangentproblem} is then solved for all combinations $k,l=1,...,d$ to obtain $\bm{q}_{kl}$ for each component of $\barbm{F}$.

{\color{mygreen}Although not utilized in this work, the sensitivities of the effective stress $\barbm{P}$ with respect to the geometrical parameters $\bm{\mu}$, which are required for applications such as shape optimization, can be computed with the geometrically parameterized formulation of the RVE as follows (in index notation)},
\begin{align}
    \frac{\partial \bar{P}_{ij}}{\partial \mu_k}
    = |\Omega^{\rm{p}}|^{-1} \int_{\Omega^{\rm{p}}} \left( A^{*\rm{p}}_{ijmn} \left(\frac{\partial}{\partial \mu_k}\left(\frac{\partial w_m^{*\text{p}}}{\partial X_r}\right)\left(F_{\bm{\mu}}^{-1}\right)_{rn} + \frac{\partial w_m^{*\text{p}}}{\partial X_r}\frac{\partial \left(F_{\bm{\mu}}^{-1}\right)_{rn}}{\partial \mu_k} \right) |\det \bm{F}_{\bm{\mu}}|+ P^{*\rm{p}}_{ij} \frac{\partial |\det \bm{F}_{\bm{\mu}}|}{\partial \mu_k} \right)d\bm{X}^{\rm{p}}. \label{eq:eff_sensitivities_full}
\end{align}
The {\color{mygreen}integrand} is complicated due to the derivatives of $\bm{F}_{\bm{\mu}}^{-1}$ and $|\det \bm{F}_{\bm{\mu}}|$, {\color{mygreen}but in principle these derivatives can be computed for a given geometrical mapping $\bm{\Phi}_{\bm{\mu}}$}. If the effective stress $\barbm{P}$ can be assumed to vary smoothly with the parameters $\bm{\mu}$, which may be a reasonable assumption for smoothly varying shapes using, for instance, splines, finite differences can be used to approximate these sensitivities.

\section{Surrogate Modelling}
\label{sec:surrogate}
Since the microscopic problem has to be solved at every macroscopic quadrature point, the solution of the microscopic PDE must be efficient. Solving it directly using FE is in general too computationally expensive, and, hence, the microscopic solver must be accelerated. In this section, a surrogate model for the geometrically parameterized microscopic PDE is developed by employing the Reduced Basis Method (RBM)~\cite{Quarteroni2015} to reduce the number of degrees of freedom and the Empirical Cubature Method (ECM)~\cite{Hernandez2017} to reduce the number of quadrature points. The key idea is to construct the surrogate model on the parent domain $\Omega^{\rm{p}}$, adapt it to each geometry $\Omega^{\bm{\mu}}$, and then solve the reduced problem.

\subsection{Reduced Basis Method}
For complex problems and geometries, typically a fine mesh is required for FE, leading to a high-dimensional solution space $\mathcal{V}_h^{\rm{p}}$ for the fluctuation displacement field $\bm{w}^{\rm{p}}$ with $\dim \mathcal{V}_h^{\rm{p}} = \mathcal{N}$. The idea of the RBM is to approximate the field with global parameter-independent basis functions and parameter-dependent coefficients, i.e.,
\begin{align}
    \bm{w}^{\rm{p}}(\bm{X}^{\rm{p}}; \barbm{F}, \bm{\mu}) \approx \sum_{n=1}^N a_n(\barbm{F}, \bm{\mu}) \bm{\phi}_n(\bm{X}^{\rm{p}}), \label{eq:pod_disp}
\end{align}
where $N$ is the number of basis functions, ideally much smaller than the dimension of the FE space, i.e., $N\ll\mathcal{N}$. The basis functions, $\{\bm{\phi}_n\}_{n=1}^N$, span a subset of $\mathcal{V}^{\rm{p}}_h$ and can be obtained by applying proper orthogonal decomposition (POD) on a set of pre-computed full solutions for different parameter values. Additionally, they are orthonormal with respect to $\mathcal{V}^{\rm{p}}$, i.e.,
\begin{align}
    (\bm{\phi}_m, \bm{\phi}_n)_{\mathcal{V}^{\rm{p}}} = \delta_{mn},
\end{align}
where $\delta_{mn}$ denotes the Kronecker delta. By utilizing the POD space for both the trial and test space and inserting $\bm{w}^{\rm{p}}$ from~\cref{eq:pod_disp} into~\cref{eq:weakform-dep-mapped,eq:Dweakform-dep}, the components for the reduced internal force vector $\mathbf{f}^{\text{POD}} \in \mathbb{R}^N$ and reduced global stiffness matrix $\mathbf{K}^{\text{POD}} \in \mathbb{R}^{N\times N}$ can be derived as
\begin{alignat}{2}
f^{\text{POD}}_i(\mathbf{a}) &\coloneqq \int_{\Omega^{\rm{p}}} \left(\frac{\partial \bm{\phi}_i}{\partial \bm{X}^{\rm{p}}} \bm{F}_{\bm{\mu}}^{-1}\right) : \bm{P}\sqbracket{\barbm{F} + \left(\sum_{n=1}^N a_n \frac{\partial \bm{\phi}_n}{\partial \bm{X}^{\rm{p}}}\right) \bm{F}_{\bm{\mu}}^{-1}} \left|\det \bm{F}_{\bm{\mu}}\right| d\bm{X}^{\rm{p}}, \label{eq:forcevector} \\
K^{\text{POD}}_{ij}(\mathbf{a}) &\coloneqq \int_{\Omega^{\rm{p}}} \left(\frac{\partial \bm{\phi}_i}{\partial \bm{X}^{\rm{p}}} \bm{F}_{\bm{\mu}}^{-1}\right) : \bm{A}\sqbracket{\barbm{F} + \left(\sum_{n=1}^N a_n \frac{\partial \bm{\phi}_n}{\partial \bm{X}^{\rm{p}}}\right) \bm{F}_{\bm{\mu}}^{-1}} : \left(\frac{\partial \bm{\phi}_j}{\partial \bm{X}^{\rm{p}}} \bm{F}_{\bm{\mu}}^{-1}\right) \left|\det \bm{F}_{\bm{\mu}}\right| d\bm{X}^{\rm{p}},
\label{eq:stiffnessmatrix}
\end{alignat}
where $\mathbf{a}=[a_1,\dots,a_N]^T$ is the column vector of unknown coefficients to be solved for, and $i, j=1,\dots, N$ span over all basis functions. Analogously to \cref{eq:full_system,eq:full_newton}, the resulting non-linear system of equations
\begin{align}
    \mathbf{f}^{\text{POD}}(\mathbf{a}) = \bm{0} \label{eq:red_system}
\end{align}
can be solved using Newton method:
\begin{align}
\begin{aligned}
\mathbf{K}^{\text{POD}}(\mathbf{a}^m)\Delta\mathbf{a} &= -\mathbf{f}^{\text{POD}}(\mathbf{a}^m), \\
    \mathbf{a}^{m+1} &= \mathbf{a}^m + \Delta\mathbf{a}.
\end{aligned} \label{eq:red_newton}
\end{align}

\subsection{Empirical Cubature Method}
\label{subsec:ECM}
Even though the solution field and linear system of equations have been reduced to dimension $N\ll\mathcal{N}$, computing the components of the force vector in~\cref{eq:forcevector} and global stiffness matrix in~\cref{eq:stiffnessmatrix} still requires integrating over the RVE. For the full integration, a numerical quadrature rule (usually based on Gauss quadrature) with integration points and corresponding weights $\{(\hatbm{X}_q, \hat{w}_q)\}_{q=1}^{\hat{Q}}$, where $\hat{Q}$ is the total number of integration points, is employed, i.e.,
\begin{align}
    f^{\text{POD}}_i(\mathbf{a}) \approx \sum_{q=1}^{\hat{Q}} \hat{w}_q \left.\left[\left(\frac{\partial \bm{\phi}_i}{\partial \bm{X}^{\rm{p}}} \bm{F}_{\bm{\mu}}^{-1}\right) : \bm{P}\sqbracket{\barbm{F} + \left(\sum_{n=1}^N a_n \frac{\partial \bm{\phi}_n}{\partial \bm{X}^{\rm{p}}}\right) \bm{F}_{\bm{\mu}}^{-1}} \left|\det \bm{F}_{\bm{\mu}}\right|\right]\right|_{\hatbm{X}_q}, \label{eq:forcevector_qp}
\end{align}
for $i=1,\dots, N$. For a fine mesh, $\hat{Q}$ is very large and thus evaluating~\cref{eq:forcevector_qp} leads to high computational costs. To address this issue, we employ the Empirical Cubature Method (ECM), which was proposed in Hern\'{a}ndez et al.~\cite{Hernandez2017} for a fixed geometry, and extend it to parameterized geometries.

The idea of ECM is to find a subset of points $\{\bm{X}_q\}_{q=1}^Q \subset \{\hatbm{X}_q\}_{q=1}^{\hat{Q}}$ with $Q\ll\hat{Q}$ among the set of all integration points with corresponding weights $\{w_q\}_{q=1}^Q$ that approximates~\cref{eq:forcevector_qp} up to a user-defined error $\varepsilon$. To find such a subset that approximates~\cref{eq:forcevector_qp} well for all admissible geometrical parameters $\bm{\mu}$,~\cref{eq:forcevector_qp} is first rewritten as
\begin{align}
\begin{aligned}
    f^{\text{POD}}_i(\mathbf{a}) &= \sum_{q=1}^{\hat{Q}} \hat{w}_q \left.\left[ \frac{\partial \bm{\phi}_i}{\partial \bm{X}^{\rm{p}}} : \underbrace{\left(\bm{P}\sqbracket{\barbm{F} + \left(\sum_{n=1}^N a_n \frac{\partial \bm{\phi}_n}{\partial \bm{X}^{\rm{p}}}\right) \bm{F}_{\bm{\mu}}^{-1}} \bm{F}_{\bm{\mu}}^{-T} \left|\det \bm{F}_{\bm{\mu}}\right|\right)}_{\bm{W}\coloneqq} \right]\right|_{\hatbm{X}_q}\\
    &=\sum_{q=1}^{\hat{Q}} \hat{w}_q \left.\left[ \frac{\partial \bm{\phi}_i}{\partial \bm{X}^{\rm{p}}} : \bm{W}(\bm{X}^{\rm{p}}; \barbm{F}, \bm{\mu}) \right]\right|_{\hatbm{X}_q},
\end{aligned}
\label{eq:forcevector_rewritten}
\end{align}
where the weighted stress $\bm{W}$ is defined. To remove the parameter dependence of the integrand in~\cref{eq:forcevector_rewritten}, the weighted stress is approximated by another reduced basis, i.e.,
\begin{align}
    \bm{W}(\bm{X}^{\rm{p}}; \barbm{F}, \bm{\mu}) \approx \sum_{l=1}^L \alpha_l(\barbm{F}, \bm{\mu}) \bm{B}_l(\bm{X}^{\rm{p}}), \label{eq:weighted_stress_pod}
\end{align}
where $\{\bm{B}_l\}_{l=1}^L$ is a set of $L$ basis functions obtained using POD, which are orthonormal with respect to $L^2(\Omega)$, i.e.,
\begin{align}
    \int_{\Omega} \bm{B}_m : \bm{B}_n d\bm{X}^{\rm{p}} = \delta_{mn}. \label{eq:pod_P_orthonormal}
\end{align}
Inserting~\cref{eq:weighted_stress_pod} into~\cref{eq:forcevector_rewritten} and rearranging yields,
\begin{align}
    f^{\text{POD}}_i(\mathbf{a}) \approx \sum_{l=1}^L \alpha_l(\barbm{F},\bm{\mu}) \sum_{q=1}^{\hat{Q}} \hat{w}_q \left.\left[\frac{\partial \bm{\phi}_i}{\partial \bm{X}} : \bm{B}_l\right]\right|_{\hatbm{X}_q}, \qquad i=1,\dots, N. \label{eq:integralsbasis}
\end{align}
Since~\cref{eq:integralsbasis} should be accurate for any choice of coefficients $\alpha_l(\barbm{F},\bm{\mu})$, all the $N\cdot L$ terms in~\cref{eq:integralsbasis} that approximate the integral have to be approximated as accurately as possible. Hence, the goal becomes to find a subset $Q(\ll\hat{Q})$ of integration points with corresponding weights $\{(\bm{X}_q, w_q)\}_{q=1}^Q$ that approximates~\cref{eq:integralsbasis} well, i.e.,
\begin{align}
\sum_{q=1}^{\hat{Q}} \hat{w}_q \left.\left[\frac{\partial \bm{\phi}_i}{\partial \bm{X}} : \bm{B}_l\right]\right|_{\hatbm{X}_q} \approx \sum_{q=1}^Q w_q \left.\left[ \frac{\partial \bm{\phi}_i}{\partial \bm{X}} : \bm{B}_l \right] \right|_{\bm{X}_q}, \qquad i=1,\dots, N,\ l=1,\dots, L.\label{eq:integral_ecm}
\end{align}
These $Q$ points and corresponding weights are found using a greedy algorithm, the details of which can be found in~\cite{Hernandez2017} and are omitted here. The algorithm is terminated when the mean squared error of all $N\cdot L$ terms is less than a user-defined tolerance $\varepsilon$.

Compared to the original algorithm for a fixed geometry, as proposed in~\cite{Hernandez2017}, the only differences are that the weighted stress $\bm{W}$ is employed instead of the stress $\bm{P}$ and that the parent domain $\Omega^{\rm{p}}$ is considered instead of a fixed domain $\Omega$. With the ECM integration rule, the hyper-reduced force vector and global stiffness matrix are computed as
\begin{alignat}{2}
f^{\text{PODECM}}_i(\mathbf{a}) &\coloneqq \sum_{q=1}^Q w_q \left.\left[\left(\frac{\partial \bm{\phi}_i}{\partial \bm{X}^{\rm{p}}} \bm{F}_{\bm{\mu}}^{-1}\right) : \bm{P}\sqbracket{\barbm{F} + \left(\sum_{n=1}^N a_n \frac{\partial \bm{\phi}_n}{\partial \bm{X}^{\rm{p}}}\right) \bm{F}_{\bm{\mu}}^{-1}} \left|\det \bm{F}_{\bm{\mu}}\right|\right]\right|_{\bm{X}_q}, \label{eq:forcevector_ecm} \\
K^{\text{PODECM}}_{ij}(\mathbf{a}) &\coloneqq \sum_{q=1}^Q w_q \left.\left[\left(\frac{\partial \bm{\phi}_i}{\partial \bm{X}^{\rm{p}}} \bm{F}_{\bm{\mu}}^{-1}\right) : \bm{A}\sqbracket{\barbm{F} + \left(\sum_{n=1}^N a_n \frac{\partial \bm{\phi}_n}{\partial \bm{X}^{\rm{p}}}\right) \bm{F}_{\bm{\mu}}^{-1}} : \left(\frac{\partial \bm{\phi}_j}{\partial \bm{X}^{\rm{p}}} \bm{F}_{\bm{\mu}}^{-1}\right) \left|\det \bm{F}_{\bm{\mu}}\right|\right]\right|_{\bm{X}_q}.
\label{eq:stiffnessmatrix_ecm}
\end{alignat}

{\color{mygreen}
\begin{remark}
    The computational costs of the ECM greedy algorithm as proposed in~\cite{Hernandez2017} increase drastically with the number of selected integration points, since for every selected point a non-negative least squares problem needs to be solved. As pointed out in~\cite{Chapman2017AcceleratedModels}, rank-one updates can be used with the least squares solver for better efficiency, and such a refined version of the ECM algorithm was presented in~\cite{Hernandez2020AECM-hyperreduction}. For the numerical examples considered in this work, the original ECM algorithm in~\cite{Hernandez2017} was sufficiently fast and we did not use the algorithmically improved version.
\end{remark}
}

\subsection{Effective Quantities}\label{subsec:eff_quant}
Once the new set of integration points and weights is found, the integrands of~\cref{eq:forcevector_ecm,eq:stiffnessmatrix_ecm} only need to be evaluated at the points $\{\bm{X}_q\}_{q=1}^Q$ during the solution of the reduced problem. This also means that the stress and stiffness field are available at these points only. To compute the effective quantities, the most straightforward method is to use the integration rule obtained by ECM, i.e.,
\begin{align}
\begin{aligned}
    \barbm{P} &= |\Omega^{\rm{p}}|^{-1}\int_{\Omega^{\rm{p}}} \bm{P}^{*\rm{p}} |\det \bm{F}_{\bm{\mu}}| d\bm{X}^{\rm{p}} \\
    &\approx |\Omega^{\rm{p}}|^{-1} \sum_{q=1}^Q w_q \left.\left(\bm{P}^{*\rm{p}} |\det \bm{F}_{\bm{\mu}}|\right)\right|_{\bm{X}_q}.
\end{aligned} \label{eq:eff_stress_1}
\end{align}
Since the stress field $\bm{P}^{*\rm{p}}$ is known at all integration points $\{\bm{X}_q\}_{q=1}^Q$, the effective stress can be directly evaluated. The method yields very accurate results in the examples considered below in~\cref{sec:results}. However, it should be noted that there is currently no guarantee that the integration rule found by ECM will generally be accurate for the computation of the effective stress. {\color{mygreen} In general, the effective stress can have two sources of error as compared to the full solution: one comes from the solution of the reduced system and one from an inaccurate integration of the obtained stress field. The ECM integration points are selected such that the first error is minimized, but this also indirectly affects the second one to decrease, although not as quickly. This can be observed in the results of the first numerical example presented in~\cref{subsec:ex1}. To ensure an accurate integration of the effective stress, it could be included into the ECM algorithm as a criterion}.

As discussed in~\cref{subsec:eff_quantities_full}, derivatives $\dfrac{\partial \bm{w}^{*\text{p}}}{\partial \barbm{F}}$ are needed to find the effective stiffness $\barbm{A}$, see~\cref{eq:eff_stiffness_full}. For each component of $\barbm{F}$, the linear tangent problem of~\cref{eq:tangentproblem} needs to be solved. By employing the trial space of the fluctuation field for the auxiliary function $\bm{q}_{kl}$, i.e.,
\begin{align}
    \bm{q}_{kl} = \sum_{n=1}^N q_n \bm{\phi}_n(\bm{X}^{\rm{p}}),
\end{align}
and the integration rule found by ECM, the following linear system of equations results:
\begin{align}
    \mathbf{K}^{*\rm{p}} \mathbf{q} = \mathbf{b}, \label{eq:tangent_problem_red}
\end{align}
where $\mathbf{q} = [q_1,\dots,q_N]^T$ is the column vector of unknowns to be solved for and
\begin{align}
    K^{*\rm{p}}_{ij} &= \sum_{q=1}^Q w_q \left.\left[ \left(\frac{\partial \bm{\phi}_i}{\partial \bm{X}^{\rm{p}}} \bm{F}_{\bm{\mu}}^{-1}\right) : \bm{A}^{*\rm{p}} : \left(\frac{\partial \bm{\phi}_j}{\partial \bm{X}^{\rm{p}}} \bm{F}_{\bm{\mu}}^{-1}\right) \left|\det \bm{F}_{\bm{\mu}}\right| \right]\right|_{\bm{X}_q}, \\
    b_i &= -\left( \sum_{q=1}^Q w_q \left.\left[ \left(\frac{\partial \bm{\phi}_i}{\partial \bm{X}^{\rm{p}}} \bm{F}_{\bm{\mu}}^{-1}\right) : \bm{A}^{*\rm{p}} \left|\det \bm{F}_{\bm{\mu}}\right|\right]\right|_{\bm{X}_q} \right) : \mathbb{E}_{kl}.
\end{align}
Note that the matrix $\mathbf{K}^{*\rm{p}}\in\mathbb{R}^{N\times N}$ is exactly the same as the global stiffness matrix $\mathbf{K}$ of~\cref{eq:stiffnessmatrix_ecm} evaluated at the solution $\bm{w}^{*\rm{p}}$. After solving the tangent problems, the effective stiffness $\barbm{A}$ can be computed (in index notation) as
\begin{align}
    \bar{A}_{ijkl} &= |\Omega^{\rm{p}}|^{-1} \sum_{q=1}^Q w_q \left.\left(\frac{\partial P_{ij}^{*\rm{p}}}{\partial \bar{F}_{kl}} |\det \bm{F}_{\bm{\mu}}|\right)\right|_{\bm{X}_q}, \label{eq:eff_stiffness_1}
\end{align}
where
\begin{align}
\begin{aligned}
    \frac{\partial \bm{P}^{*\rm{p}}}{\partial \bar{F}_{kl}} &= \bm{A}^{*\rm{p}} : \left( \mathbb{E}_{kl} + \left(\frac{\partial \bm{q}_{kl}}{\partial \bm{X}^{\rm{p}}} \bm{F}_{\bm{\mu}}^{-1} \right)\right). \label{eq:eff_stress_f11}
\end{aligned}
\end{align}

\subsection{Auxiliary Problem For Geometrical Transformation}
\label{subsec:auxiliary}
Thus far, the geometrical transformation $\bm{\Phi}_{\bm{\mu}} : \Omega^{\rm{p}} \rightarrow \Omega^{\bm{\mu}}$ has been assumed to be known and has not been discussed in more detail. However, such transformations are in general not known analytically and have to be found numerically by using, for example, radial basis functions, see, e.g.,~\cite{Rozza2008, Stabile2020EfficientMethods}, or mesh-based methods, see, e.g.,~\cite{yao1989,Guo2022LearningParameterizations}. For each of those methods, an auxiliary problem arises which needs to be solved. In order to rapidly solve the surrogate model for a wide range of different geometries, it must therefore also be ensured that the auxiliary problem can be solved rapidly. In this work, the method in~\cite{Guo2022LearningParameterizations} is employed, in which the auxiliary problem is formulated as a linear elasticity problem by defining $\bm{\Phi}_{\bm{\mu}}(\bm{X}^{\rm{p}}) = \bm{X}^{\rm{p}} + \bm{d}(\bm{X}^{\rm{p}})$, with $\bm{d}$ the transformation displacement obtained from
\begin{alignat}{2}
    \text{Div} \left(\mathbb{C}^{\text{aux}} : \frac{1}{2}\left( \frac{\partial\bm{d}}{\partial\bm{X}^{\rm{p}}} + \left(\frac{\partial\bm{d}}{\partial\bm{X}^{\rm{p}}}\right)^T\right)\right) &= \bm{0}         &&\text{ in } \Omega^{\rm{p}}. \label{eq:auxiliary_problem}
\end{alignat}
In the above equation, $\mathbb{C}^{\text{aux}}$ is the fourth-order elasticity tensor, fully specified by the Young's modulus $E^{\text{aux}}$ and Poisson's ratio $\nu^{\text{aux}}$. The boundary conditions for this PDE are problem-dependent and are specified by the geometrical parameters $\bm{\mu}$. For the RVE problem, the outer boundaries are fixed ($\bm{d}=\bm{0}$), while $\bm{d}$ is prescribed on parts of the interior that are parameterized by $\bm{\mu}$. In~\cite{Guo2022LearningParameterizations} the effect of the choice of $E^{\text{aux}}$ and $\nu^{\text{aux}}$ was studied and it was demonstrated empirically that the choice only has a minor effect on the final approximation quality. Hence, in all numerical examples considered in this work, a Young's modulus of $E^{\text{aux}}=1$ and Poisson's ratio $\nu^{\text{aux}}=0.25$ is assumed. The auxiliary problem can then be significantly accelerated with the RBM in combination with a (D)EIM~{\color{mygreen}\cite{Barrault2004,Chaturantabut2010}}, resulting in
\begin{align}
    \hatmathbf{A} \hatmathbf{d} = \hatmathbf{b}(\bm{\mu}), \label{eq:reduced_aux_problem}
\end{align}
where $\hatmathbf{A} \in \mathbb{R}^{N_p\times N_p}$ is the reduced system matrix, $\hatmathbf{d} \in \mathbb{R}^{N_p}$ is the reduced transformation displacement, $\hatmathbf{b}(\bm{\mu})\in \mathbb{R}^{N_p}$ is the reduced forcing vector and $N_p$ is the number of geometrical parameters. Since $N_p$ is usually small,~\cref{eq:reduced_aux_problem} can be rapidly solved. From $\hatmathbf{d}$, the transformation gradient $\bm{F}_{\bm{\mu}}$, its inverse $\bm{F}_{\bm{\mu}}^{-1}$, and its determinant $\det \bm{F}_{\bm{\mu}}$ can be computed. Moreover, expressions for the derivative of the inverse and determinant of the transformation gradient $\bm{F}_{\bm{\mu}}$ can be derived, which are needed for computing the sensitivities with respect to $\bm{\mu}$, see~\cref{eq:eff_sensitivities_full}. For more information on the auxiliary problem and its reduction, the reader is referred to~\cite{Guo2022LearningParameterizations}.

\subsection{Summary}
For convenience, the offline-online decomposition for constructing and solving the surrogate model is summarized in Algorithm~\ref{alg:podgpr}.

\begin{algorithm}[H]
\caption{Offline-online decomposition of the proposed PODECM framework with microstructures parameterized with external loading $\barbm{F}$ and geometrical features $\bm{\mu}$.}\label{alg:podgpr}
\begin{algorithmic}[1]
\Require
\State Define a parent domain $\Omega^{\rm{p}}$ and its finite element discretization.
\State Generate parameter samples $\{\barbm{F}^i,\bm{\mu}^i\}_{i=1}^{N_s}$ from a random distribution.
\State For each different set of geometrical parameters $\bm{\mu}^i$, solve the auxiliary problem in~\cref{eq:auxiliary_problem} to obtain the transformation map $\bm{\Phi}_{\bm{\mu}^i}$.
\State Compute $\bm{F}_{\bm{\mu}^i}^{-1}$ and $\det \bm{F}_{\bm{\mu}^i}$ for each parameter sample $\bm{\mu}^i$, then run full simulations (\cref{eq:weakform-dep-mapped,eq:Dweakform-dep}) for $\barbm{F}^i$ and collect fluctuation displacement and weighted stress snapshots.
\State Compute POD for the fluctuation displacement and weighted stress, cf.~\cref{eq:pod_disp,eq:weighted_stress_pod}.
\State Run ECM algorithm and find integration points and weights, cf.~\cref{eq:integral_ecm}.
\State Assemble the reduced system matrix and forcing vector for the auxiliary problem in~\cref{eq:reduced_aux_problem} by applying POD and DEIM. Details are provided in~\cite{Guo2022LearningParameterizations}.
\Ensure
\setcounter{ALG@line}{0}
\State Given a new parameter set $(\barbm{F}^*,\bm{\mu}^*)$, solve reduced auxiliary problem~\cref{eq:reduced_aux_problem} and compute $\bm{F}_{\bm{\mu}^*}^{-1}$ and $\det \bm{F}_{\bm{\mu}^*}$.
\State Solve reduced problem for $\barbm{F}^*$ with~\cref{eq:forcevector_ecm,eq:stiffnessmatrix_ecm}.
\State Compute effective stress using~\cref{eq:eff_stress_1}.
\State Solve the linear problem~\cref{eq:tangent_problem_red} for each component of $\barbm{F}^*$.
\State Compute components of the effective stiffness with~\cref{eq:eff_stiffness_1}.
\end{algorithmic}
\end{algorithm}

\section{Example Problems}
\label{sec:results}
The proposed framework, referred to as PODECM, is first tested on a non-linear composite microstructure under various loading conditions and analyzed in depth regarding its capabilities and accuracy. The RVE consists of an elasto-plastic matrix with stiff inclusions of variable size and is considered under non-monotonic loading. The surrogate model is analyzed in terms of the number of basis functions of the fluctuation displacement field $N$, number of basis functions of the weighted stress $L$ and the ECM integration error tolerance $\varepsilon$. Subsequently, a two-scale problem involving a porous microstructure under non-monotonic loading conditions and varying porosities is studied to illustrate the accuracy and speed-up of PODECM in a two-scale setting.

All experiments are defined in two dimensions under plane strain conditions. The RVEs are assumed to be of size $[0, 1]^2$ and all quantities are assumed to be normalized and hence dimensionless. Since the macroscopic deformation gradient $\barbm{F}$ can always be decomposed into a rotation $\barbm{R}$ and a symmetric stretch tensor $\barbm{U}$ with a polar decomposition, i.e., $\barbm{F}=\barbm{R}\barbm{U}$, it is sufficient to generate training data for the stretch tensor $\barbm{U}$, having only 3 independent components (6 in 3D).

To measure the quality of the approximation, the following error measures to compare the full FE simulations against PODECM solutions are defined:
\begin{enumerate}
    \item Error of effective stress
    \begin{align}
        \epsilon_{\barbm{P}} = \frac{\sum_{k=1}^K ||\barbm{P}^{\text{PODECM}}(\barbm{U}^k) - \barbm{P}^{\text{FE}}(\barbm{U}^k)||_F}{\sum_{k=1}^K ||\barbm{P}^{\text{FE}}(\barbm{U}^k)||_F},
    \end{align}
    where $\barbm{P}^{\text{PODECM}}(\barbm{U}^k)$ and $\barbm{P}^{\text{FE}}(\barbm{U}^k)$ denote the effective stress obtained with PODECM and FE for $\barbm{U}^k$, $||\bullet||_F$ denotes the Frobenius norm, $K$ is the total number of loading steps and $\barbm{U}^k$ is the applied external load at load step $k$.
    \item Error of fluctuation field
    \begin{align}
        \epsilon_{\bm{w}} = \frac{\sum_{k=1}^K ||\bm{w}^{\text{PODECM}}(\barbm{U}^k) - \bm{w}^{\text{FE}}(\barbm{U}^k)||_{\mathcal{V}}}{\sum_{k=1}^K ||\bm{w}^{\text{FE}}(\barbm{U}^k)||_{\mathcal{V}}},
    \end{align}
    where $\bm{w}^{\text{PODECM}}$ and $\bm{w}^{\text{FE}}$ denote the fluctuation displacement field obtained with PODECM and FE, and $||\bullet||_{\mathcal{V}}^2=(\bullet, \bullet)_{\mathcal{V}}$, c.f.,~\cref{eq:Vdotproduct}. Recall that the integral in~\cref{eq:Vdotproduct} is defined over the parameterized domain $\Omega^{\bm{\mu}}$.
\end{enumerate}

\subsection{Elasto-Plastic Composite RVE With Random Stiff Inclusions}\label{subsec:ex1}
\subsubsection{Problem Description}
\begin{figure}[ht]
    \centering
    \begin{subfigure}[b]{0.245\textwidth}
        \centering
        \includegraphics[width=\textwidth]{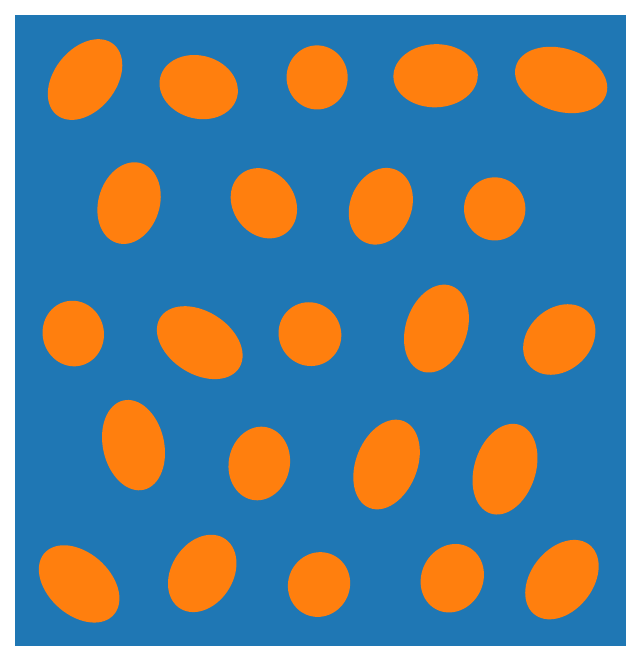}
        \caption{Parent domain $\Omega^{\rm{p}}$ with $\zeta=1$}
        \label{fig:ex1_geom_a}
    \end{subfigure}
    \begin{subfigure}[b]{0.245\textwidth}
        \centering
        \includegraphics[width=\textwidth]{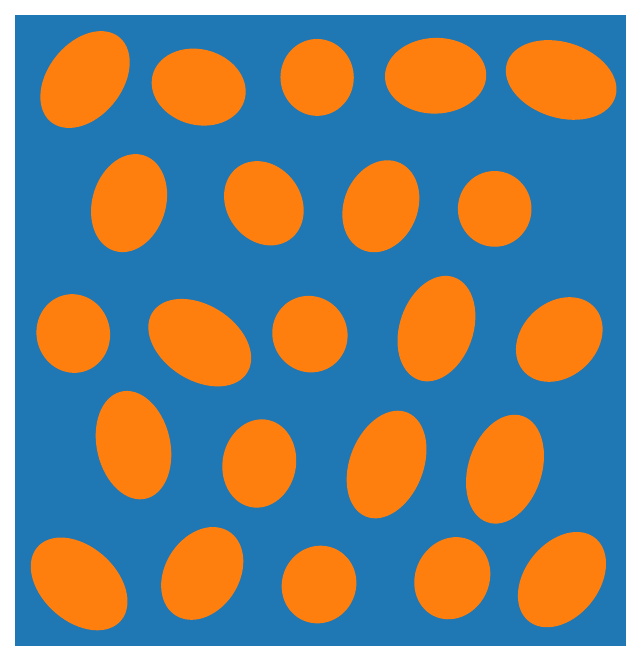}
        \caption{Domain $\Omega^{\bm{\mu}_1}$ with $\zeta=1.2$}
        \label{fig:ex1_geom_b}
    \end{subfigure}
    \begin{subfigure}[b]{0.245\textwidth}
        \centering
        \includegraphics[width=\textwidth]{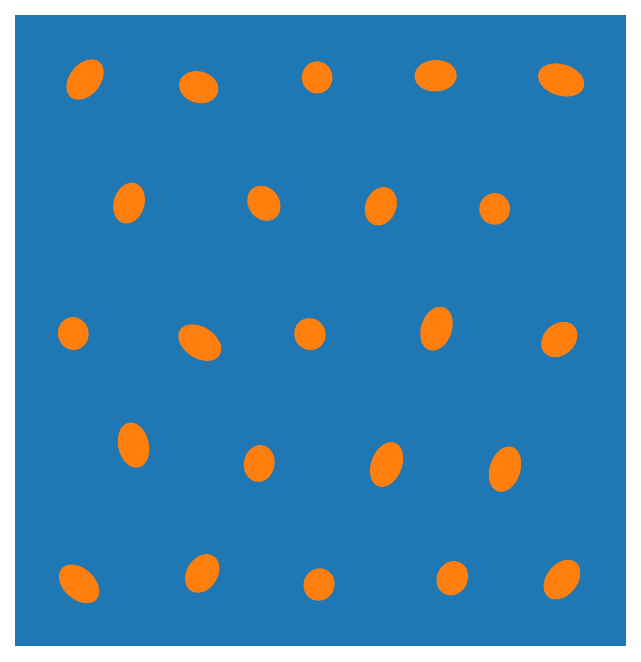}
        \caption{Domain $\Omega^{\bm{\mu}_2}$ with $\zeta=0.5$}
        \label{fig:ex1_geom_c}
    \end{subfigure}
    \begin{subfigure}[b]{0.237\textwidth}
        \centering
        \includegraphics[width=\textwidth]{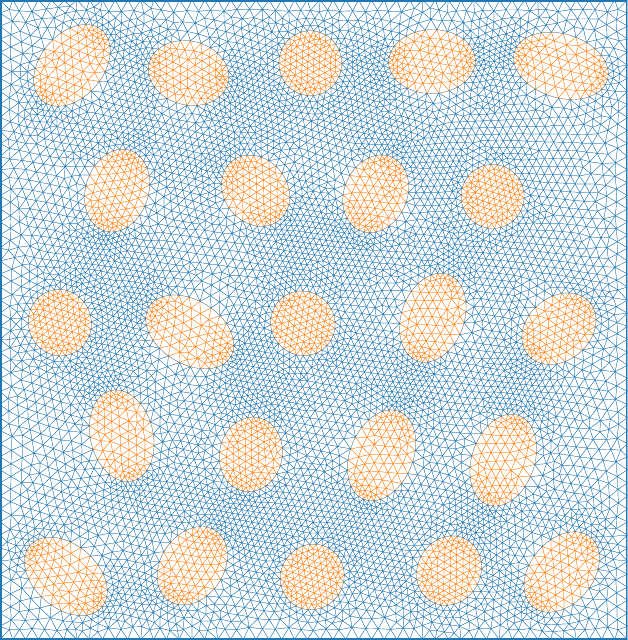}\vspace{0.25em}
        \caption{Simulation mesh}
        \label{fig:ex1_mesh}
    \end{subfigure}
    \caption{Parent with two parameterized domains and simulation mesh. (a) The parent domain consists of a matrix material (blue) with 23 random elliptical inclusions (orange). The problem has one geometrical parameter $\zeta$ that scales all ellipses uniformly ($\zeta=1$ for parent domain). (b) A parameterized domain for $\zeta=1.2$ and (c) for $\zeta=0.5$. (d) The considered mesh consists of six-noded triangular elements and contains in total 62194 degrees of freedom, 15450 triangular elements and 46350 quadrature points.}
        \label{fig:ex1_geom}
\end{figure}

The considered RVE in this example consists of two phases, an elasto-plastic matrix and stiff elastic inclusions. The geometry of the parent domain is shown in~\cref{fig:ex1_geom_a}, where the volume fraction of the inclusions is $23.4\%$. For the geometrical parameterization, one geometrical parameter $\bm{\mu}=\{\zeta\}$ that scales the size of the inclusions uniformly (and is proportional to the volume fraction of the inclusions) is introduced, see~\cref{fig:ex1_geom_b,fig:ex1_geom_c} showing two example domains for distinct values of $\zeta$. The simulation mesh is depicted in~\cref{fig:ex1_mesh}, where six-noded quadratic triangular elements are used in conjunction with three quadrature points per element. In total, the mesh has 62194 degrees of freedom, 15450 triangular elements and 46350 quadrature points.

For the constitutive model of both matrix and inclusion the small-strain $J_2$-plasticity model with linear isotropic hardening is chosen and extended to large strains with the method presented in Cuitino and Ortiz~\cite{Cuitino1992PapersKINEMATICS}. {\color{mygreen}The details of the plasticity model are provided in~\ref{appendixA}.} For the matrix, the following material parameters are selected: {\color{mygreen}a Young's modulus} $E=10$, {\color{mygreen}Poisson's ratio} $\nu=0.3$, {\color{mygreen}yield stress} $\sigma_{y0}=0.2$ and {\color{mygreen}hardening constant} $H=5$. For the inclusions, $E=100$ and $\nu=0.3$ {\color{mygreen}are} selected, corresponding to a stiffness contrast ratio of 10 {\color{mygreen}between both components}. Since no plastic deformation is assumed for the inclusions, their yield stress is set to a large value such that yielding never occurs.

\begin{figure}[ht]
    \centering
    \includegraphics[width=0.9\textwidth]{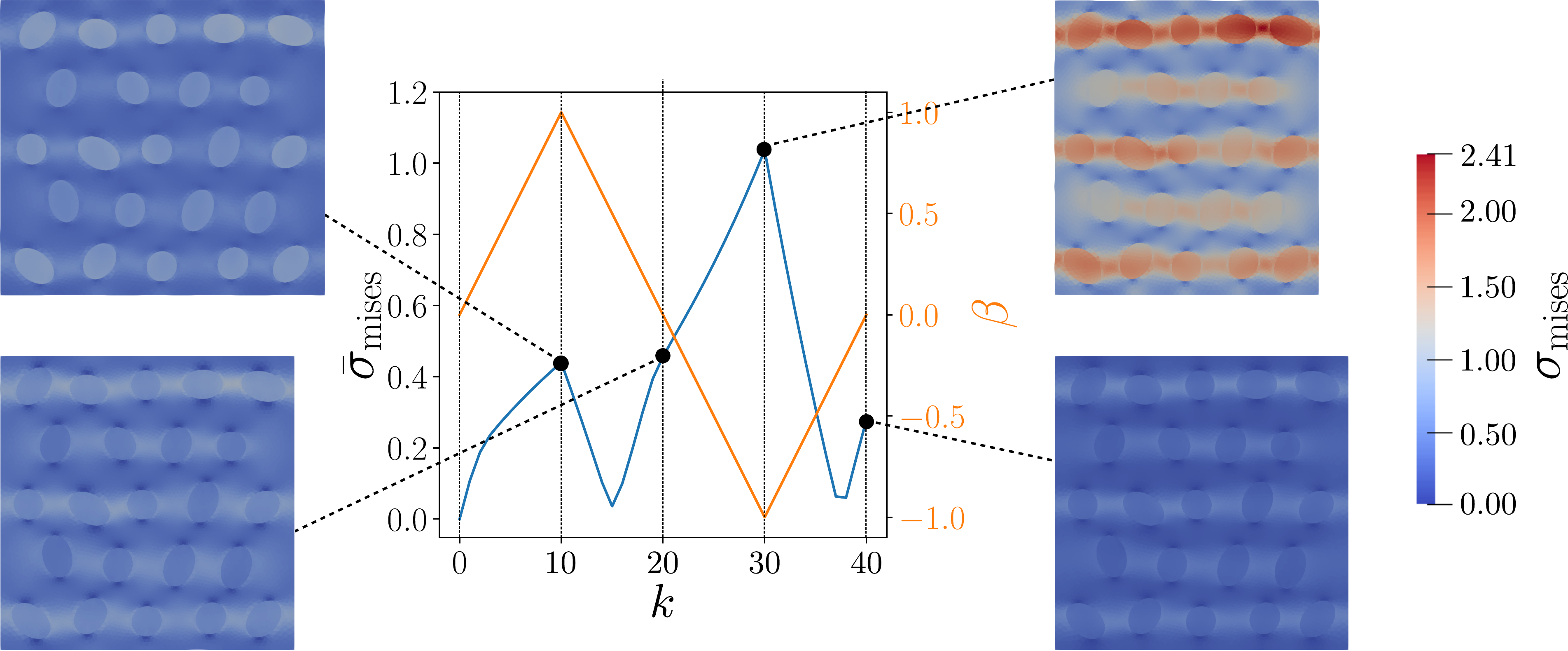}
    \caption{Macroscopic von Mises stress $\barbm{\sigma}_{\text{mises}}$ {\color{mygreen}and amplitude function $\beta$} plotted over $k$ for $\zeta=1.010$, $\bar{U}_{xx}=1.1$, $\bar{U}_{yy}=1.0$, $\bar{U}_{xy}=0.0$. At $k=\{10,20,30,40\}$, microstrucural von Mises stress fields $\bm{\sigma}_{\text{mises}}$ are shown. The von Mises stress is non-zero at $k=\{20,40\}$ due to residual plastic deformation.}
    \label{fig:ex1_example}
\end{figure}

Three loading $\bar{U}_{xx}$, $\bar{U}_{xy}$, $\bar{U}_{yy}$ and one geometrical $\zeta$ parameters are considered with bounds $\zeta\in[0.5,1.2]$, $\bar{U}_{xx}\in[0.9, 1.1]$, $\bar{U}_{yy}\in[0.9, 1.1]$ and $\bar{U}_{xy}\in[-0.1, 0.1]$. Through $\zeta$, the volume fraction of the inclusions is varied from 5.85\% to 33.7\%. {\color{mygreen}For each sample, the macroscopic right stretch tensor $\barbm{U} = \begin{bmatrix}
    \bar{U}_{xx} & \bar{U}_{xy} \\ \bar{U}_{xy} & \bar{U}_{yy}
\end{bmatrix}$ is applied to the RVE with $\beta(k)\barbm{U}$, where $\beta(k)$ is a piecewise linear amplitude function with load step $k\in\{0,1,\dots,40\}$. The chosen amplitude function is shown in~\cref{fig:ex1_example} in orange, together with the evolution of the effective von Mises stress for an example with $\zeta=1.010$, $\bar{U}_{xx}=1.1$, $\bar{U}_{yy}=1.0$, $\bar{U}_{xy}=0.0$, as well as the local von Mises stress fields at steps $k=\{10,20,30,40\}$.} Even though only macroscopic strains of up to 10\% are applied, local strains reach values up to 83\%.

\subsubsection{Results}
\begin{figure}[ht]
    \centering
    \includegraphics[width=\textwidth]{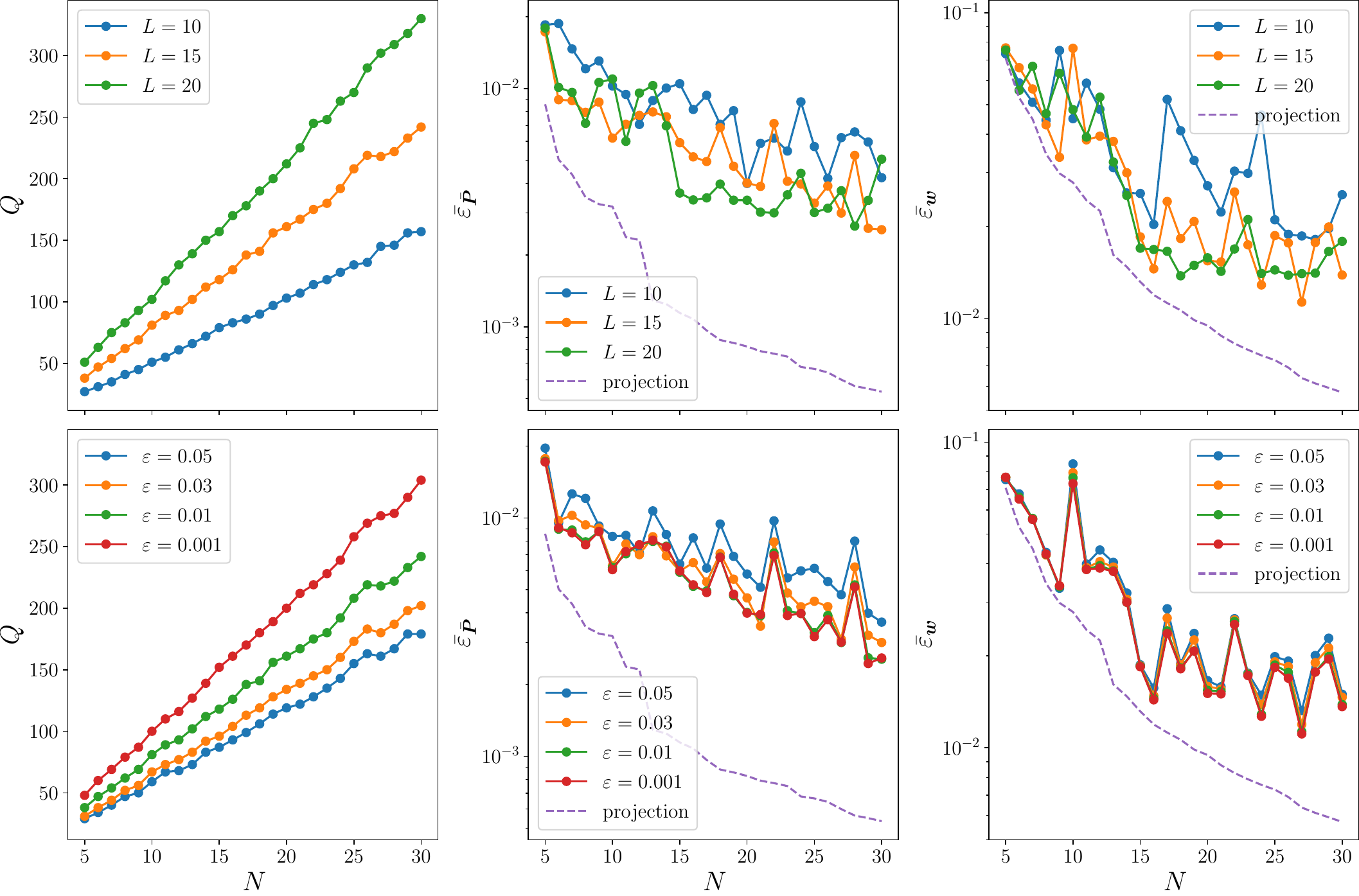}
    \caption{The left column shows the number of quadrature points $Q$ obtained from ECM for different choices of number of basis functions of fluctuation field $N$, number of basis functions of weighted stress $L$, and ECM integration error $\varepsilon$. The middle and right columns show the average errors of the effective stress and the fluctuation field when tested on the testing data for different choices of $N$, $L$ and $\varepsilon$. The top row assumes a fixed $\varepsilon=0.01$, while the bottom row assumes a fixed $L=15$.}
    \label{fig:ex1_results}
\end{figure}

\begin{figure}[ht]
    \centering
    \includegraphics[width=\textwidth]{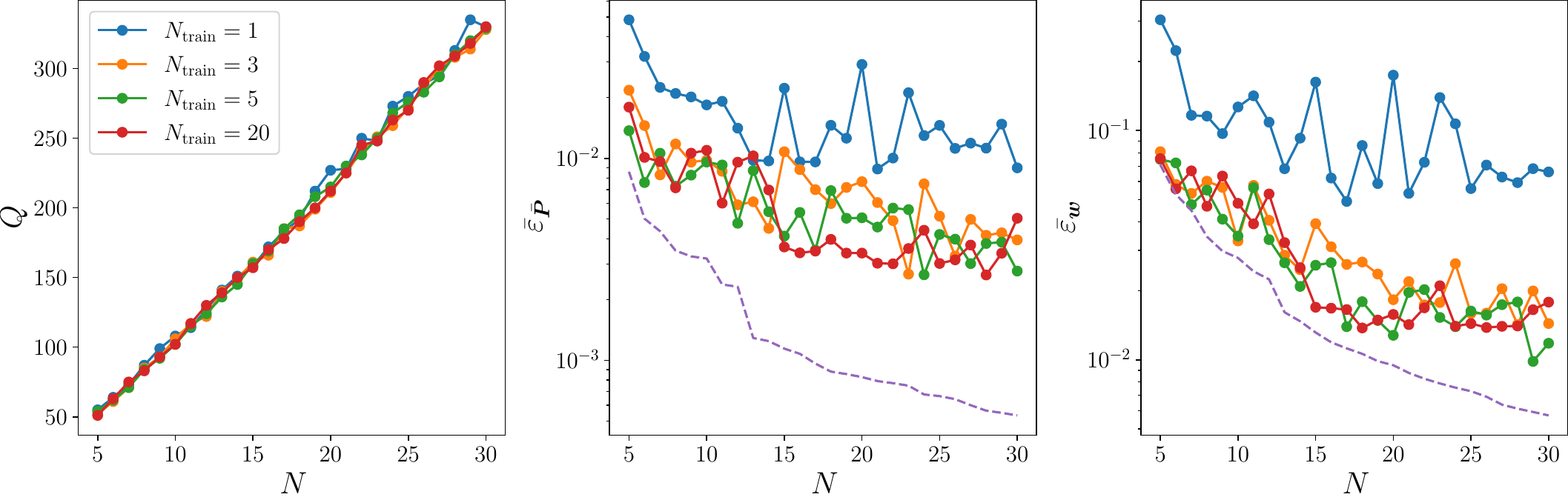}
    \caption{Errors of PODECM for different numbers of training samples. The number of quadrature points does not change. For $N_{\rm{train}}=3$, the errors in effective stress and fluctuation displacement is already converged, showing that PODECM is very data efficient.}
    \label{fig:ex1_results2}
\end{figure}

In total {\color{mygreen}$N_{\rm{train}}=20$} samples are generated from a Sobol sequence to train PODECM whereas 100 testing samples are generated from a uniform distribution to test PODECM. Each sample consists of 40 snapshots for each load step.

The accuracy and speed-up of PODECM depends on the number of basis functions used for the fluctuation displacement field $N$ and the number of quadrature points $Q$. While $N$ is typically chosen directly, $Q$ depends on the choice of the number of basis functions used for the weighted stress $L$ and the ECM integration error $\varepsilon$.

To study the influence of $L$ on the resulting number of quadrature points $Q$ and mean errors in effective stress and fluctuation field on the testing dataset, several combinations of $N$ and $L$ for a fixed $\varepsilon=10^{-2}$ are tested, with resulting errors shown in the top row of~\cref{fig:ex1_results}. The projection error (for $N$ basis functions and using full integration) is shown as well. It can be clearly seen that the number of quadrature points $Q$ increases drastically with increasing $N$ and $L$, as more information needs to be integrated accurately. {\color{mygreen}In fact, a roughly linear relationship $Q \propto NL$ can be recognized.} For the mean errors, a higher $L$ leads to better results on average, although we observe that errors fluctuate significantly, and for some values of $N$ a worse approximation is obtained with a higher $L$. This occurs since the ECM algorithm is a greedy algorithm, meaning that it does not necessarily find an optimal set of integration points. When more basis functions are included into the algorithm, a completely different set of points may be found that finally leads to a worse approximation. It can furthermore be observed that the gap between the projection error and the PODECM solution grows larger for increasing $N$. This is because the basis functions typically become more oscillatory and difficult to approximate with higher $N$, see, e.g.,~\cite{Guo2018,Guo2021b}, and thus require significantly more quadrature points for a good approximation. It is interesting that the gap for the errors in the fluctuation field are smaller than the ones in the effective stress{\color{mygreen}, i.e., the difference between the computed PODECM and the projection errors are much higher for the effective stress as compared to the fluctuation field. This happens because the ECM integration points and weights are primarily selected to integrate the weak form accurately, and using them to compute the effective stress introduces an additional approximation error, cf.~\cref{subsec:eff_quant}.}

Several combinations of $N$ and $\varepsilon$ for a fixed $L=15$ are next tested to study the influence of $\varepsilon$ on the number of quadrature points and approximation errors. Obtained results are shown in the bottom row of~\cref{fig:ex1_results}. Similarly to the previous analysis, a lower $\varepsilon$ leads to more quadrature points $Q$ and a lower mean error in the effective stress and fluctuation field on average, as the integrals are approximated more accurately. Interestingly, lowering the tolerance from 0.01 to 0.001 does not significantly improve the approximation quality, even though substantially more quadrature points are included, meaning that the errors can be attributed to the higher modes of the weighted stress (the additional quadrature points barely contain any information). Therefore, choosing a tolerance smaller than $\varepsilon=0.01$ leads to no improvement.

From~\cref{fig:ex1_results} we further observe that the errors of the fluctuation field are considerably higher (order of magnitude) than the errors of the effective stresses. This results from the fact that the POD basis functions aim to minimize the $H^1(\Omega^{\rm{p}})$ error, and thus approximate the field accurately on average rather than locally, suggesting a favorable approximation for averaged quantities such as the effective stress. In~\cite{Schneider2022SuperconvergenceMaterials}, the authors showed a similar result for FFT: the effective stresses converge with an order of $h$, while the local fields converge with $h^{1/2}$, where $h$ is the used voxel edge-length.

{\color{mygreen}To test the data efficiency of PODECM, the reduced model is trained for different numbers of training data $N_{\text{train}}=\{1, 3, 5, 20\}$ with $L=20$ and $\varepsilon=0.01$. The number of integration points and corresponding errors in the effective stress and fluctuations are shown in~\cref{fig:ex1_results2}. It can clearly be seen that the number of integration points barely changes for different $N_{\text{train}}$, and that the errors converged already for $N_{\text{train}} = 3$. No noticeable improvements with $N_{\text{train}} = 5$ and 20 can be observed, showing that PODECM is very data efficient. Even with $N_{\text{train}} = 1$, the errors in the effective stress are below 5\%.}

To conclude, the more basis functions $N$ and $L$ are used and the lower the integration error $\varepsilon$ is chosen, the more accurate the final result is. However, at the same time the surrogate model grows in size and the speed-up decreases. A user must thus make a compromise between accuracy and cost. In our experience, the speed-up correlates nearly linearly with $Q$, i.e., if the number of quadrature points is reduced by a factor of 100, this results in a speed-up of roughly 100. In contrast, the number of basis functions $N$ only plays a minor role for the speed-up. For this example, use of $N=20$, $L=20$ and $\varepsilon=0.01$ lead to a reduction in the number of degrees of freedom from 62194 to 20 and quadrature points from 46350 to 212, suggesting a speed-up on the order of roughly 200.

{\color{mygreen}
\begin{remark}
    Intuitively, one might consider not selecting $L$ as an independent parameter, but as a (non-linear) function of $N$ with $L\geq N$, since the weighted stress acts as a non-linear function on the fluctuation field, and thus is expected to require more basis functions. However, for $L=N$, one obtains a roughly quadratic relationship for the number of quadrature points $Q$ and $N$ with $Q\propto LN=N^2$. If a higher number of $N$ is necessary to have a sufficiently large solution space, $Q$ quickly becomes very large and speed-ups of PODECM become very small. By treating $L$ as an independent parameter and allowing $L<N$, the resulting number of integration points can be controlled and decreased. Furthermore, we observed in the numerical tests that, if a maximum number of integration points is specified, we often obtained better results for $N>L$ rather than $N\leq L$, as long as $L$ is large enough.
\end{remark}}

\subsection{Two-Scale Compression With Porous Microstructure}\label{subsec:ex2}
\subsubsection{Problem Description}
\begin{figure}[ht]
    \centering
    \begin{subfigure}[b]{0.37\textwidth}
        \centering
        \includegraphics[width=\textwidth]{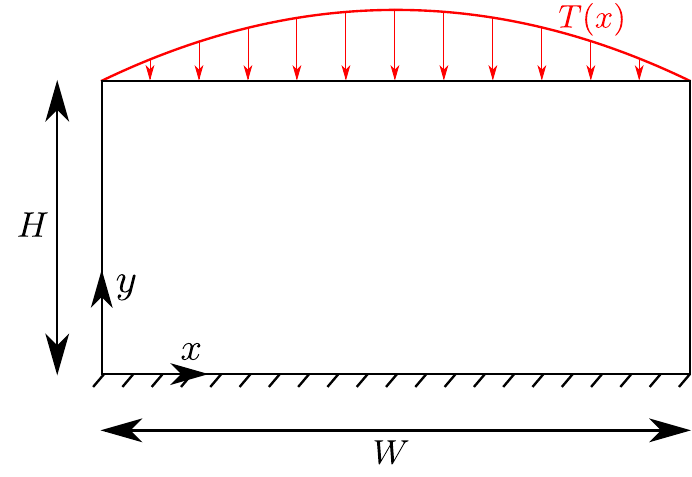}
        \caption{Macroscopic structure}
        \label{fig:ex2_geom_a}
    \end{subfigure}\hspace{2em}
    \begin{subfigure}[b]{0.33\textwidth}
        \centering
        \includegraphics[width=\textwidth]{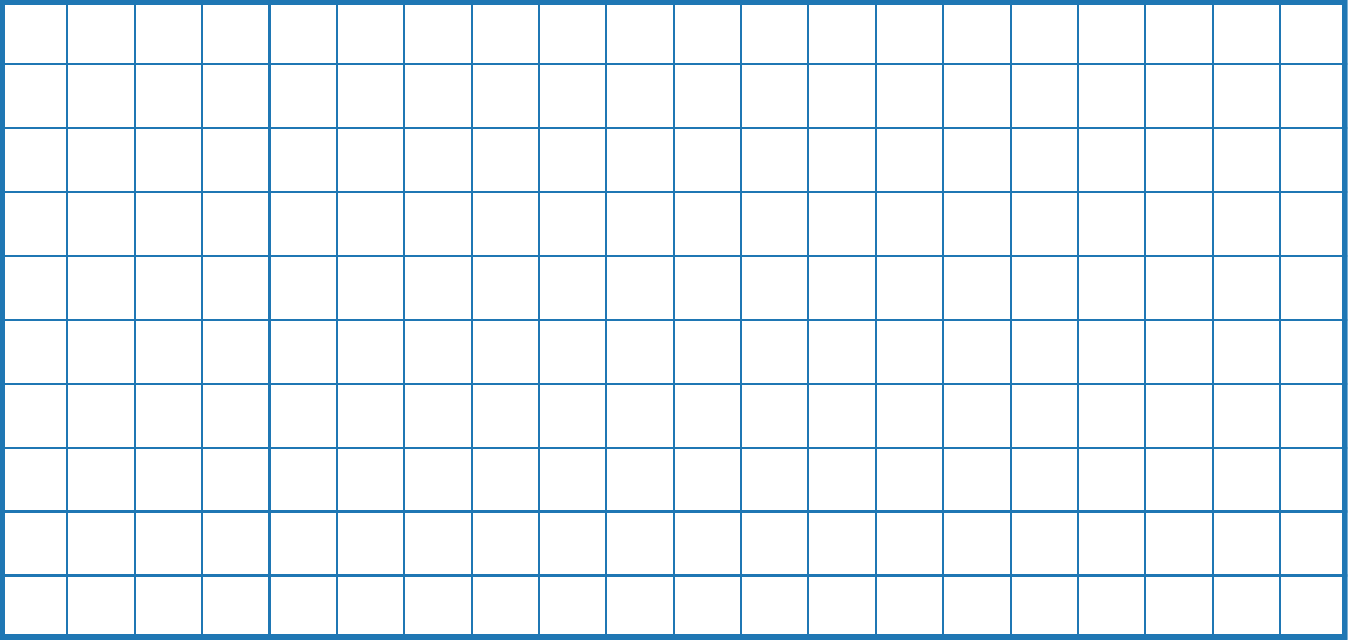}
        \vspace{1.4em}
        \caption{Macroscopic simulation mesh}
        \label{fig:ex2_geom_b}
    \end{subfigure}
    \caption{Geometry (a) and mesh (b) of the considered macroscopic structure. The body is fixed on the bottom and an external compression force $T$ is applied on the top. The mesh consists of 1322 degrees of freedom, 200 8-noded quadrilateral elements and 4 quadrature points per element.}
    \label{fig:ex2_geom}
\end{figure}

\begin{figure}[ht]
    \centering
    \begin{subfigure}[b]{0.35\textwidth}
        \centering
        \includegraphics[width=\textwidth]{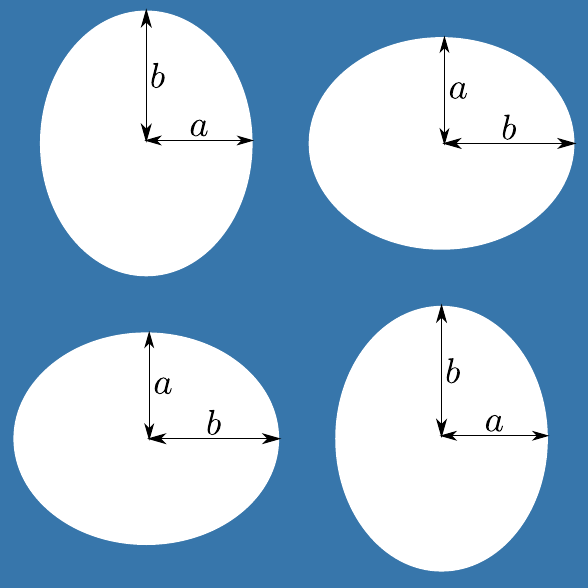}
        \caption{RVE}
        \label{fig:ex2_rve_a}
    \end{subfigure}\hspace{2em}
    \begin{subfigure}[b]{0.34\textwidth}
        \centering
        \includegraphics[width=\textwidth]{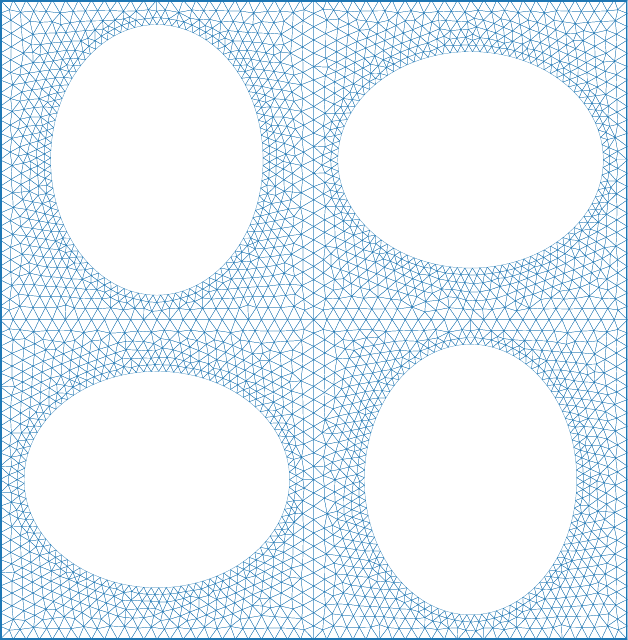}
        \caption{Simulation mesh}
        \label{fig:ex2_rve_b}
    \end{subfigure}
    \caption{Geometry (a) and mesh (b) of the porous RVE. The elliptical holes {\color{mygreen}are all characterized by the same} semi-minor axis $a$ and semi-major axis $b$, and are parameterized by the volume fraction of the pores $v_{\text{void}}$ and the $\kappa=b/a$. The employed simulation mesh has 21042 degrees of freedom, 4964 6-noded triangular elements and in total 14892 quadrature points. The parent domain corresponds to $\kappa=1.25$ and $v_{\text{void}}=0.45$.}
    \label{fig:ex2_rve}
\end{figure}

In the second example, the macroscopic structure, depicted in~\cref{fig:ex2_geom} together with the employed simulation mesh, is compressed under an external loading $T(x)$. Here, we assume $H=1$, $W=2$, $T(x)=\bar{T}\left(1-\left(\dfrac{2x}{W}-1\right)^2\right)$, $x\in[0,W]$, with $\bar{T}$ the magnitude of the applied load. The simulation mesh has 1322 degrees of freedom, 200 8-noded quadrilateral elements, and in total 800 quadrature points. The structure is assumed to have a porous microstructure, modelled by the parameterized RVE, shown in~\cref{fig:ex2_rve}, and the same material model as the matrix material in the previous example. Such microstructures (with circular holes, i.e., $a=b$) have been considered in several works, see, e.g.,~\cite{Bertoldi2008MechanicsStructures,Zhang2021AMaterials,Ameen2018SizeTransformations,Rokos2019a}, due to their auxetic behavior under compression, i.e., negative Poisson's ratio. During compression, the center part of the material starts to rotate, thus pulling the material from the sides inwards. In our work, we define two independent parameters, namely the volume fraction of the voids, $v_{\text{void}}\coloneqq 4\pi a b$, and the ratio of the semi-major axis $b$ to semi-minor axis $a$ of each hole, $\kappa\coloneqq b/a$. The semi-minor axis $a$ and semi-major axis $b$ depend on $v_{\text{void}}$ and $\kappa$ as
\begin{align}
    a(v_{\text{void}}, \kappa) &= \sqrt{v_{\text{void}} / (4 \pi \kappa)},\\
    b(v_{\text{void}}, \kappa) &= \kappa \cdot a(v_{\text{void}}, \kappa).
\end{align}

\begin{figure}[ht]
    \centering
    \begin{subfigure}[b]{0.41\textwidth}
        \centering
        \includegraphics[width=\textwidth]{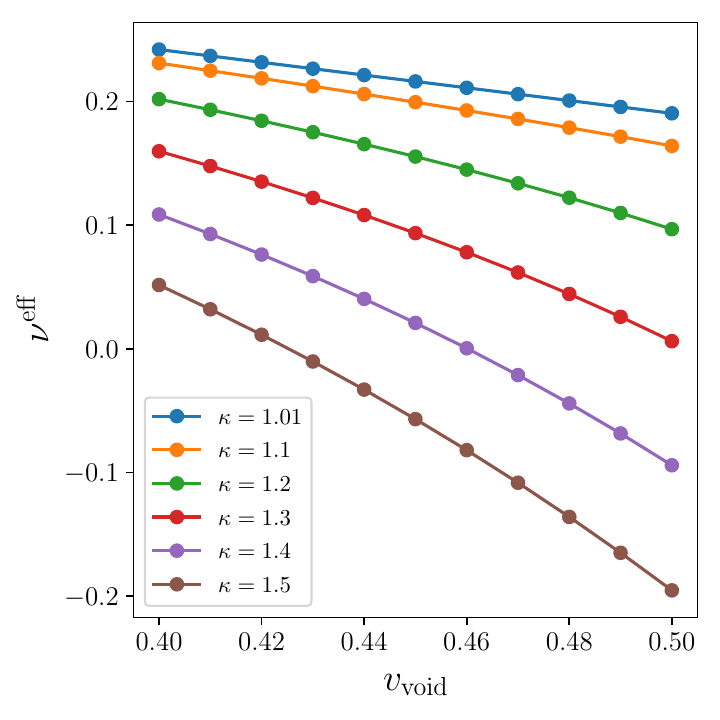}
        \caption{Effective Poisson's ratio $\nu^{\text{eff}}$}
        \label{fig:ex2_rve_params_a}
    \end{subfigure}\hspace{1em}
    \begin{subfigure}[b]{0.41\textwidth}
        \centering
        \includegraphics[width=\textwidth]{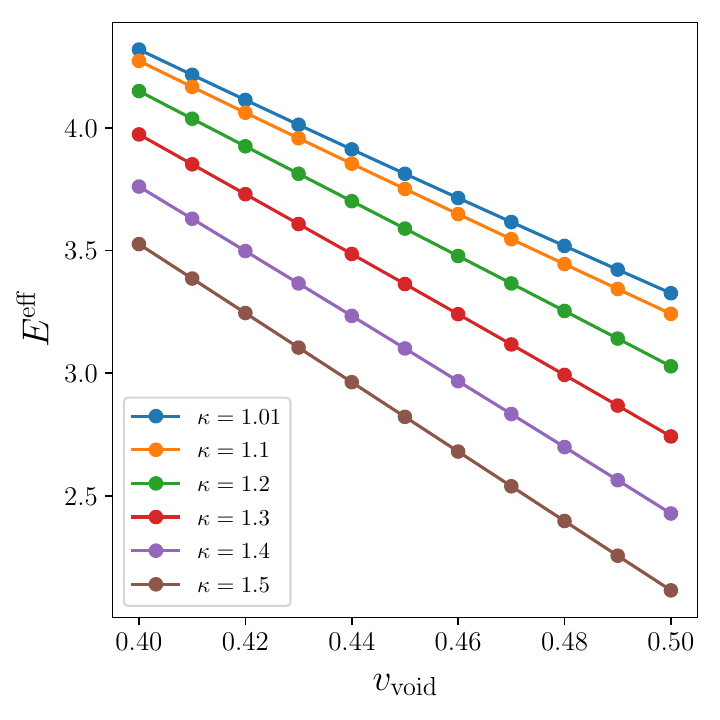}
        \caption{Effective Young's modulus $E^{\text{eff}}$}
        \label{fig:ex2_rve_params_b}
    \end{subfigure}
    \caption{Initial effective Poisson's ratio (a) and Young's modulus (b) of RVE for different values of $v_{\text{void}}$ and $\kappa$.}
    \label{fig:ex2_rve_params}
\end{figure}

Depending on the values of the parameters, the resulting effective properties change significantly. To illustrate this, linear analyses of this RVE for different parameters have been carried out, similarly to \cite{Zhang2022ProgrammableMetamaterials}, where a small compression in the $y$-direction with $\Delta u_y=0.001$ has been applied, while allowing the RVE to contract freely in the $x$-direction. With the resulting displacement in the $x$-direction, $\Delta u_x$, the Poisson's ratio in the initial state can be estimated as
\begin{align}
    \nu^{\text{eff}} = -\frac{\Delta u_x}{\Delta u_y}.
\end{align}
Similarly, the initial Young's modulus is estimated as
\begin{align}
    E^{\text{eff}} = \frac{\bar{P}_{yy}}{\Delta u_y},
\end{align}
where $\bar{P}_{yy}$ is the $yy$-component of the effective stress. For parameter ranges $v_{\text{void}}\in [0.4, 0.5]$ and $\kappa\in [1.01, 1.5]$, the estimated Poisson's ratio and Young's modulus are plotted in~\cref{fig:ex2_rve_params}. It can be observed that removing material (by increasing $v_{\text{void}}$ while keeping $\kappa$ fixed) or increasing $\kappa$ while keeping $v_{\text{void}}$ fixed both lead to a softer response with lower Young's modulus. While the Poisson's ratio is barely affected by $v_{\text{void}}$ for values of $\kappa$ close to 1, the effect becomes apparent for larger values of $\kappa$. In particular, for $\kappa\geq1.4$, the Poisson's ratio changes from a positive value to a negative one. Therefore, by tuning $v_{\text{void}}$ and $\kappa$, the RVE behavior can be significantly modified.

The parameters $v_{\text{void}}$ and $\kappa$ are chosen to vary smoothly through the macrostructural domain as
\begin{align}
    \kappa(x, y) &= 1.5 - (1.5-1.01)y, \\
    v_{\text{void}}(x, y) &= 0.4 + (0.5 - 0.4) (1 - x) ^ 2.
\end{align}
The parent mesh is selected with $\kappa=1.25$ and $v_{\text{void}}=0.45$. The external loading is applied in $K=50$ load steps. In the first 25 load steps, the applied load $\bar{T}$ increases linearly from 0 to 0.2. In the next 25 steps $\bar{T}$ is decreased linearly from 0.2 to 0.

\subsubsection{Results}
Several two-scale simulations with different PODECM surrogate models are run and compared to the full reference FE$^2$ solution. To compare the accuracy of the surrogate solutions, the compliance $C \coloneqq \int_{\Gamma} T(x) u_y(x) dx$, where $\Gamma$ denotes the top horizontal edge of the macrostructure and $u_y$ its vertical displacement, is computed at every load step. The compliance is an important quantity, often employed in optimization problems. Subsequently, the relative error in compliance $\epsilon_C$ and the relative error averaged over all load steps $\bar{\epsilon}_C$ are defined as
\begin{align}
    \epsilon_{C,k} \coloneqq \frac{|C_k-C_k^{\text{FE}^2}|}{|C_k^{\text{FE}^2}|}, \qquad
    \bar{\epsilon}_C \coloneqq \frac{1}{K} \sum_{n=1}^{K} \epsilon_{C,k},
\end{align}
where the subscript $k$ denotes the $k$-th load step and $C^{\text{FE}^2}$ is the compliance computed with the full solution.

The tested PODECM RVE models are generated for different numbers of training samples $N_{\text{train}}$ and number of basis functions $N$. The training data is sampled from $\bar{U}_{xx}\in[0.85, 1]$, $\bar{U}_{yy}\in[0.85, 1]$, $\bar{U}_{xy}\in[-0.15,0.15]$, $v_{\text{void}}\in [0.4, 0.5]$ and $\kappa\in [1.01, 1.5]$ with a Sobol sequence. Each sample consists of 40 load steps, where the {\color{mygreen}sampled macroscopic stretch tensor is applied to the RVE with a piecewise linear amplitude function that is linearly increased from 0 to 1 for the first 20 load steps and then linearly decreased from 1 to 0 for the last 20 steps}. For the ECM algorithm, the number of basis functions for the weighted stress $L$ and the integration error $\varepsilon$ are assumed as fixed with $L=20$ and $\varepsilon=0.01$ for all models. The exact settings for each surrogate model are summarized in \cref{table:ex2_comparison_roms}, alongside the averaged relative error in compliance $\bar{\epsilon}_C$, and the run time and {\color{mygreen}online} speed-up in comparison to the full FE$^2$ solution. For comparison, the total number of degrees of freedom and quadrature points of the full FE model of the RVE are also provided. As can be seen, for all surrogate models the number of quadrature points are reduced by a factor of up to 100 which results in high speed-ups up to 100 times, while errors are below 5\% for all models. By increasing $N$ from 10 to 50 for rom\_1 to rom\_5, the error decreases from 4.74\% to 1.54\%, whereas the speed-up reduces from 92 to 26. {\color{mygreen}Including more training samples with the same number of basis functions $N=50$ (from $N_{\text{train}}=20$ for rom\_5 to $N_{\text{train}}=50$ for rom\_6) improves the error from 1.54\% to 0.39\% while the speed-up remains roughly the same. This means that by increasing the sample size, the first 50 basis functions contain more general information that result in a better approximation. When more than 50 samples are used for rom\_7 and rom\_8, the error remains roughly the same, implying that 50 samples are sufficient for this problem and including more training data does not improve the results.}

\begin{table}[ht]
\centering
\caption{Summary of results for full FE$^2$ and different PODECM surrogate models. Reduced order models are generated for different numbers of training samples $N_{\text{train}}$ and basis functions of the displacement $N$. The number of quadrature points $Q$ follow from the ECM algorithm with a fixed number of basis functions of the weighted stress field $L=20$ and an integration error of $\varepsilon=0.01$. All reduced order models achieve errors less than 5\% with speed-ups up to 100 times. By generating more training data and maintaining the same $N$ (rom\_5 and rom\_6), better results are achieved. {\color{mygreen}However, using more than 50 training samples (rom\_7 and rom\_8) leads to no further improvements, suggesting that 50 samples are sufficient for this problem.}}
\begin{tabular}{c|ccc|ccc}
       & $N_{\text{train}}$ & $N$   & $Q$   & $\bar{\epsilon}_C$ & run time\footnote{All computations are executed using 20 cores of an Intel Platinum 8260.} & {\color{mygreen}online} speed-up \\ \hline
full   & -                  & 21042 & 14892 & -            & 12573s    & -        \\
rom\_1 & 20                 & 10    & 132   & 4.74\%       & 133s     & 94.53    \\
rom\_2 & 20                 & 20    & 259   & 4.66\%       & 195s     & 64.48    \\
rom\_3 & 20                 & 30    & 372   & 2.28\%       & 260s     & 48.36    \\
rom\_4 & 20                 & 40    & 490   & 2.19\%       & 337s     & 37.31    \\
rom\_5 & 20                 & 50    & 595   & 1.54\%       & 438s     & 28.71    \\
rom\_6 & 50                 & 50    & 591   & 0.39\%       & 426s     & 29.51    \\
rom\_7 & 80                 & 50    & 579   & 0.58\%       & 422s     & 29.79    \\
rom\_8 & 100                & 50    & 577   & 0.65\%       & 411s     & 30.59
\end{tabular}
\label{table:ex2_comparison_roms}
\end{table}

\begin{figure}[ht]
    \centering
    \includegraphics[width=0.75\textwidth]{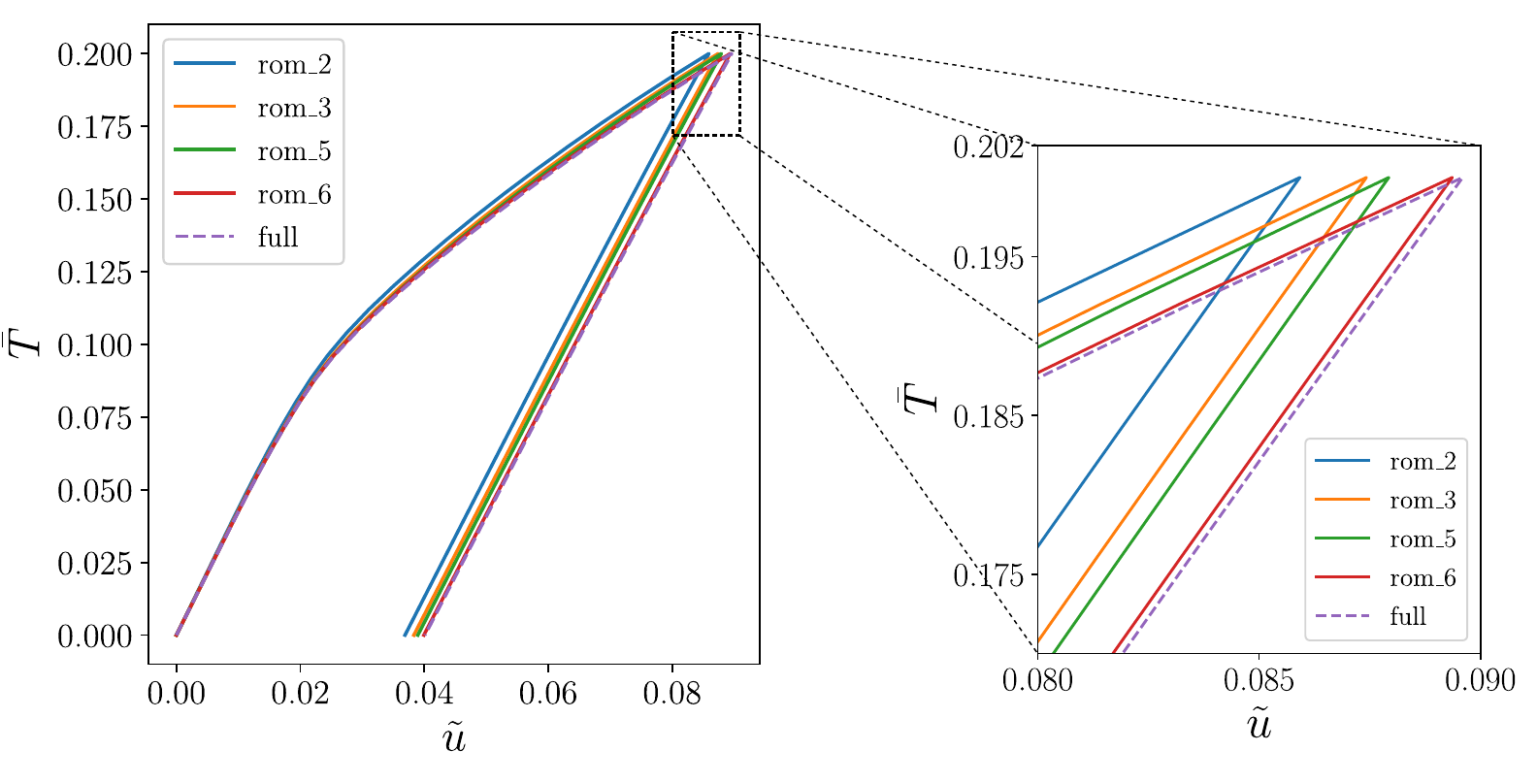}
    \caption{Force-displacement curve of two-scale simulations. The displacement $\tilde{u}$ is defined as the vertical displacement of the mid point on the top edge, cf.~\cref{fig:ex2_geom_a}. The structure starts deforming plastically for $\bar{T}>0.07$ and a residual displacement of roughly $\tilde{u}=0.04$ remains after unloading. All surrogate models achieve accurate results, although they generally predict a slightly stiffer response than the full FE$^2$ solution.}
    \label{fig:ex2_force_disp}
\end{figure}

In~\cref{fig:ex2_force_disp} the force-displacement curve is shown for the FE$^2$ and a few selected surrogate solutions. The displacement is defined as the vertical displacement at the mid point of the top edge (which is also the maximal displacement). It can be observed that all surrogate models underpredict the displacement, indicating that the surrogate models overpredict the stiffness of the macrostructure.

\begin{figure}[ht]
    \centering
    \includegraphics[width=0.7\textwidth]{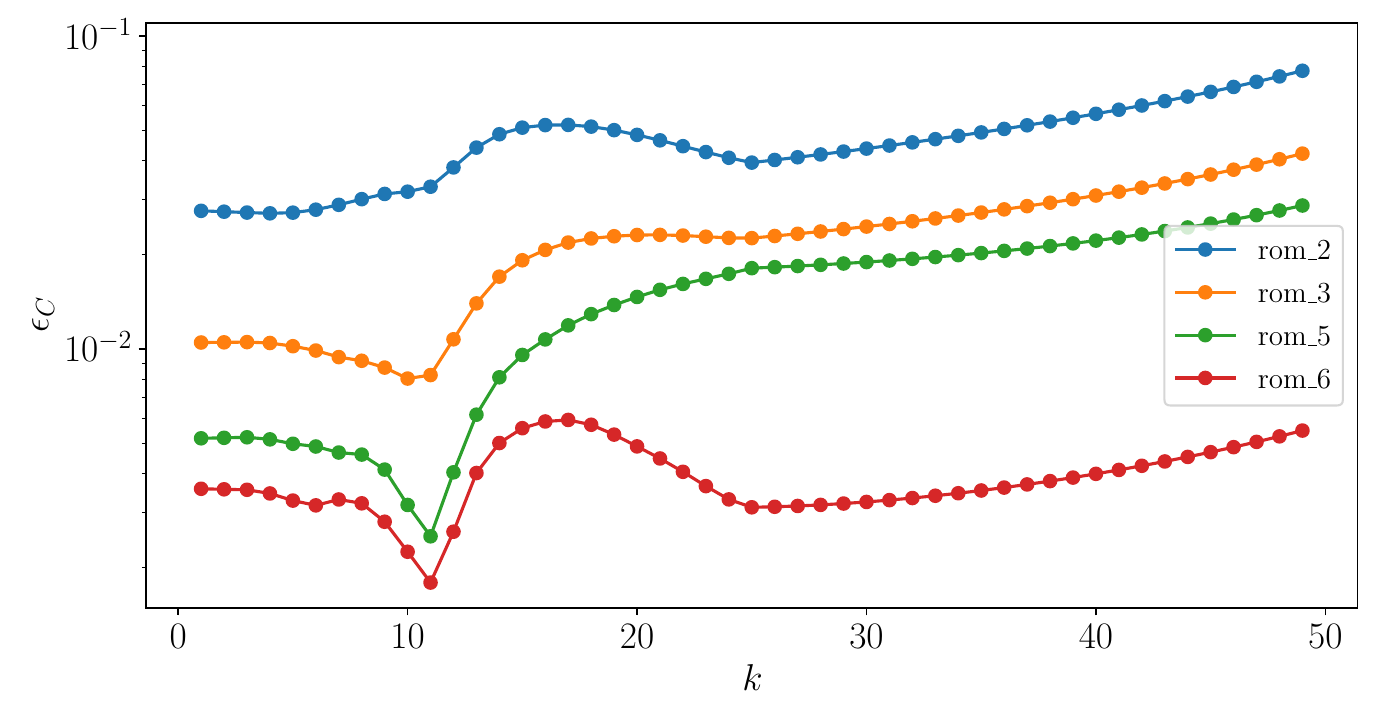}
    \caption{Relative error $\epsilon_C$ in compliance over load step $k$. All surrogate models (specified in~\cref{table:ex2_comparison_roms}) begin with a lower error which slowly grows with $k$. By increasing the sample size of the training data (comparing rom\_5 and rom\_6), the prediction becomes more accurate for the same number of basis functions; the errors for all load steps for rom\_6 are below 1\%.}
    \label{fig:ex2_compliance}
\end{figure}

The relative error in compliance is plotted over the load steps $k$ in~\cref{fig:ex2_compliance}.
For all models, the error slowly increases over $k$. The reason for this behavior is that all training samples are generated for simple loading cases, where the macroscopic stretch tensor is linearly varied in only one direction throughout the entire simulation. In the macroscopic simulation, the macroscopic stretch tensor for one integration point generally does not evolve along one direction, but changes its direction continuously, leading to highly complicated deformation paths and histories that are not included in the training data. To tackle this problem, random loading paths during training could be used, as performed in, e.g.,~\cite{Mozaffar2019,Wu2020a}, to generate a more general surrogate model. Solely increasing the number of samples from 20 to {\color{mygreen}50} (rom\_5 to rom\_6) also decreases the observed errors to less than 1\% for all load steps. This shows that PODECM generalizes well to loading paths that are not part of the training data.

{\color{mygreen}Regarding the offline run times (for one core of an Intel Platinum 8260): computing each training sample takes approximately 2--3 minutes. With the training data available, computing the POD for each model takes up to around one minute. Finally, selecting the ECM points takes up to roughly two minutes for the PODECM models with $N=50$.}

\section{Conclusions}
\label{sec:conclusion}
In this work, we developed a reduced order model, termed PODECM, by combining the proper orthogonal decomposition (POD), the empirical cubature method (ECM) and a geometrical transformation method. This method is designed to accelerate microscopic computations within two-scale simulations, parameterized by material and geometrical parameters, hence enabling PODECM to be used for two-scale optimization problems. We showed how to compute the effective stress and the corresponding consistent effective stiffness for this model, as required by the macroscopic solver, and how to obtain the effective sensitivities with respect to geometrical parameters. The framework was first tested on a single-scale problem involving an RVE of a composite microstructure that consisted of a soft elasto-plastic matrix with stiff inclusions of variable size, controlled by a single geometrical parameter. With PODECM, the number of degrees of freedom and integration points was reduced to a fraction of the full FE model while maintaining a high accuracy in effective stress. The performance of PODECM was further evaluated for a two-scale simulation, in which a porous microstructure, characterized by two geometrical parameters, was considered. Both geometrical parameters were varied throughout the macrostructure, and depending on their values, the effective Poisson's ratio changed from positive to negative. For this example, different PODECM models were constructed with good accuracies (of errors less than 1\%), while achieving speed-ups up to 100.
Even though highly accurate and fast solutions were obtained with the proposed method, several open questions prevail. For instance, optimality and accuracy of the ECM integration rule cannot be ensured and thus needs to be further analyzed and understood.

{\color{mygreen}As the underlying microscopic PDE is still being solved, only a small amount of training data is required to construct a good approximation of the full model. History-dependent material behavior does not need to be specially treated, making the proposed framework very general and viable for two-scale shape optimization problems.}

\section*{Data availability}
The data that support the findings of this study are available from the corresponding author upon request.
\section*{Acknowledgements}
This result is part of a project that has received funding from the European Research Council (ERC) under the European Union’s Horizon 2020 Research and Innovation Programme (Grant Agreement No. 818473). The authors would in addition like to thank Martin Hor\'{a}k from Czech Technical University in Prague for his help with the implementation of the large-strain $J_2$-plasticity model.
\bibliography{references}
\appendix
\section{Plasticity model of the RVE}\label{appendixA}
The small-strain $J_2$-plasticity model with linear isotropic hardening model obeys:
\begin{align}
    \bm{\sigma} &= \mathbb{D} : (\bm{\epsilon} - \bm{\epsilon}^{\color{mygreen}\text{pl}}), \label{eq:j2_1}\\
    f^{\text{yield}} &= ||\bm{\sigma}||_{\text{mises}} - (\sigma_{y0} + H\xi),\label{eq:j2_2}\\
    \bm{r} &= \frac{\partial f}{\partial \bm{\sigma}} = \sqrt{\frac{3}{2}} \frac{\operatorname{Dev}(\bm{\sigma})}{\sqrt{\operatorname{Dev}(\bm{\sigma}):\operatorname{Dev}(\bm{\sigma})}},\label{eq:j2_3}\\
    \dot{\xi} &= \gamma, \label{eq:j2_4}\\
    \dot{\bm{\epsilon}}^{\color{mygreen}\text{pl}} &= \gamma \bm{r}, \label{eq:j2_5}\\
    \gamma &\geq 0, \ f^{\text{yield}} \leq 0, \ \gamma f^{\text{yield}}=0, \label{eq:j2_6}
\end{align}
where $\bm{\epsilon}$ is the small-strain tensor, $\bm{\epsilon}^{\color{mygreen}\text{pl}}$ the plastic strain tensor, $\mathbb{D}$ is the fourth-order elasticity tensor that can be fully specified by Young's modulus $E$ and Poisson's ratio $\nu$, and $\bm{\sigma}$ is the corresponding stress tensor; $f^{\text{yield}}$ defines the yield surface, $\operatorname{Dev}(\bullet)$ takes the deviatoric part of a tensor $(\bullet)$, $||\bullet||_{\text{mises}}=\sqrt{\dfrac{3}{2}\operatorname{Dev}(\bullet) : \operatorname{Dev}(\bullet)}$ computes the von Mises stress, $H$ is the hardening constant, $\sigma_{y0}$ yield stress, $\bm{r}$ the plastic flow direction, $\xi$ the equivalent plastic strain that defines the isotropic hardening of the yield surface, and $\gamma$ is the consistency parameter. {\color{mygreen}The dot above a quantity denotes the time derivative, which is typically approximated with an implicit Euler scheme with some discretized time step. However, since we consider rate-independent plasticity here, the time step can always be multiplied with the $\gamma$ and does not play a role. Instead, time steps become load steps, and sometimes very small load steps are required for the global Newton solver to converge. Given the last converged step $k$ with internal variables $(\bm{\epsilon}^{pl,k}, \xi^k)$ and a new input strain $\bm{\epsilon}^{k+1}$, the new internal variables are obtained with $\bm{\epsilon}^{{pl,k+1}}=\bm{\epsilon}^{{pl,k}}+\gamma \bm{r}^{k+1}$ and $\xi^{k+1}=\xi^k+\gamma$, where $\bm{r}^{k+1}$ is the new plastic flow direction, computed from the new stress $\bm{\sigma}^{k+1}$. The value for the consistency parameter $\gamma$ is computed such that the Kuhn-Tucker conditions~\cref{eq:j2_6} are fulfilled. $\xi$ and $\bm{\epsilon}^{\text{pl}}$ are both assumed to be $0$ and $\bm{0}$ at the start, i.e., $\xi^0=0$ and $\bm{\epsilon}^{{pl,0}}=0$.} For more information on the material model see Simo and Hughes~\cite{Simo2006}.

Following the procedure of Cuitino et al.~\cite{Cuitino1992PapersKINEMATICS}, by employing a multiplicative split of the deformation gradient $\bm{F} = \bm{F}^{\text{el}} \bm{F}^{\text{pl}}$, {\color{mygreen}where $\bm{F}$ is split into its elastic $\bm{F}^{\text{el}}$ and plastic $\bm{F}^{\text{pl}}$ part with $\bm{F}^{\text{pl}}=\bm{I}$ at the beginning}, the elastic logarithmic strain can be defined as
\begin{align}
    \bm{C}^{\text{el}}_{\text{log}} \coloneqq \ln{\bm{C}^{\text{el}}} = \ln((\bm{F}^{\text{el}})^T \bm{F}^{\text{el}}). \label{eq:j2_7}
\end{align}
By interpreting the elastic logarithmic strain ${\color{mygreen}\bm{C}^{\text{el}}_{\text{log}}}$ as the small-strain tensor ${\color{mygreen}\bm{\epsilon}}$, the small-strain constitutive model defined in~\cref{eq:j2_1,eq:j2_2,eq:j2_3,eq:j2_4,eq:j2_5,eq:j2_6} is used to compute the stress tensor $\hatbm{S}$ on the intermediate configuration,
\begin{align}
    \hatbm{S} \coloneqq 2 \bm{\sigma}(\bm{C}^{\text{el}}_{\text{log}}) : \deriv{\bm{C}^{\text{el}}_{\text{log}}}{\bm{C}^{\text{el}}}, \label{eq:j2_8}
\end{align}
while the 1PK stress is recovered from $\hatbm{S}$ as
\begin{align}
    \bm{P} = (\bm{F}^{\text{el}})^{-1} \hatbm{S} \ (\bm{F}^{\rm{pl}})^{-T}. \label{eq:j2_9}
\end{align}
Instead of evolving the plastic strain with~\cref{eq:j2_5}, the plastic deformation gradient $\bm{F}^{\rm{pl}}$ is evolved according to {\color{mygreen}the following incremental form
\begin{align}
    \bm{F}^{{\rm{pl},k+1}} = \exp(\gamma\bm{r}^{k+1})\bm{F}^{{\rm{pl},k}}, \label{eq:j2_10}
\end{align}}
{\color{mygreen}where $\bm{F}^{{\rm{pl},0}}=\bm{I}$.}
\end{document}